\pdfoutput=1

\RequirePackage{etex}

\documentclass[pdftex,colorlinks,notitlepage,a4paper,12pt]{article}

\usepackage{slantsc}

\usepackage[a4paper, margin=1in]{geometry}

\usepackage[pdftex]{graphicx}

\usepackage{epstopdf}

\usepackage[usenames,dvipsnames]{color}
\usepackage[dvipsnames]{xcolor}

\usepackage{soul}
\soulregister\parencite7
\soulregister\textcite7
\soulregister\cite7
\soulregister\posscite7
\soulregister\ref7
\soulregister\enquote7

\newlength{\mylength}
\makeatletter
\newcommand{\mycfs}[1]{%
	\normalsize
	\@defaultunits\mylength=#1pt\relax\@nnil
	\edef\@tempa{{\strip@pt\mylength}}%
	\ifx\protect\@typeset@protect
	\edef\@currsize{\noexpand\mycfs\@tempa}% store calculated size
	\fi
	\mylength=1.2\mylength
	\edef\@tempa{\@tempa{\strip@pt\mylength}}%
	\expandafter\fontsize\@tempa
	\selectfont
}

\usepackage[utf8]{inputenc}
\usepackage[T1]{fontenc}

\usepackage[french, german, spanish, british]{babel}
\babeltags{es = spanish}
\babeltags{de = german}
\usepackage[style=american]{csquotes}

\usepackage[en-GB]{datetime2}
\DTMlangsetup[en-GB]{ord=raise}

\usepackage{textgreek}

\usepackage{latexsym}

\usepackage{amsmath}

\usepackage{econometrics}

\usepackage{amssymb}

\usepackage{bbold}

\usepackage{pdflscape} % \begin{landscape} ... \end{landscape}

\usepackage{afterpage}

\usepackage{array}
\usepackage{delarray}
\usepackage{cases}
\usepackage{rotating}
\usepackage{longtable}
\usepackage{threeparttable}
\usepackage{threeparttablex}
\usepackage{booktabs}
\usepackage{caption}
\usepackage{subcaption}
\usepackage{makecell}

\usepackage{tabularx}
\usepackage{ltablex}
\keepXColumns
\usepackage{tabu}
\usepackage{linegoal}
\usepackage{multirow}

\makeatletter
\newlength\LongtableWidth
\newcommand*{\org@longtable}{}
\let\org@longtable\longtable
\def\longtable{%
	\begingroup
	\advance\c@LT@tables\@ne
	\edef\x{LT@\romannumeral\c@LT@tables}
	\global\LongtableWidth\z@
	\@ifundefined{\x}{%
	}{%
		\def\LT@entry##1##2{%
			\global\advance\LongtableWidth##2\relax
		}%
		\@nameuse{\x}%
	}%
	\typeout{* \x: \the\LongtableWidth}%
	\endgroup \ifdim\LongtableWidth>\z@ \setlength{\LTcapwidth}{\LongtableWidth}%
	\fi
	\org@longtable }
\makeatother

\usepackage{ragged2e}

\usepackage{xpatch}

\usepackage{float}
\usepackage{placeins}

\usepackage{eurosym}    % \euro{} and \euro{1}

\usepackage{appendix}

\usepackage[hyphens]{url}

\usepackage{xurl}

\hyphenchar\font=\string"7F
\usepackage[super]{nth}

\usepackage[shortlabels]{enumitem} % a package for nice enumerations
\setlist[enumerate]{nosep, label = {(\arabic*)}}
\setlist[itemize]{nosep, label = {---}}

\usepackage{titlesec}
\titlelabel{\thetitle.\quad}
\titleformat*{\section}{\normalfont\large\bfseries\setstretch{1.5}}
\titleformat*{\subsection}{\normalfont\large\bfseries\setstretch{1}}

\usepackage{fancyhdr}

\makeatletter
\newread\pin@file
\newcounter{pinlineno}
\newcommand\pin@accu{}
\newcommand\pin@ext{pintmp}
\newcommand*\partialinput [3] {%
	\IfFileExists{#3}{%
		\openin\pin@file #3
		\setcounter{pinlineno}{1}
		\@whilenum\value{pinlineno}<#1 \do{%
			\read\pin@file to\pin@line
			\stepcounter{pinlineno}%
		}
		\addtocounter{pinlineno}{-1}
		\let\pin@accu\empty
		\begingroup
		\endlinechar\newlinechar
		\@whilenum\value{pinlineno}<#2 \do{%
			\readline\pin@file to\pin@line
			\edef\pin@accu{\pin@accu\pin@line}%
			\stepcounter{pinlineno}%
		}
		\closein\pin@file
		\expandafter\endgroup
		\scantokens\expandafter{\pin@accu}%
	}{%
		\errmessage{File `#3' doesn't exist!}%
	}%
}
\makeatother

\usepackage[backend=biber,style=apa,uniquelist=false,uniquename=false,dashed=true,dateabbrev=false,eventdate=comp,hyperref=true,language=british,parentracker=true]{biblatex}

\DeclareNameFormat{labelname:poss}{% Based on labelname from biblatex.def
	\nameparts{#1}% Not needed if using Biblatex 3.4
	\ifcase\value{uniquename}%
	\usebibmacro{name:family}{\namepartfamily}{\namepartgiven}{\namepartprefix}{\namepartsuffix}%
	\or
	\ifuseprefix
	{\usebibmacro{name:first-last}{\namepartfamily}{\namepartgiveni}{\namepartprefix}{\namepartsuffixi}}
	{\usebibmacro{name:first-last}{\namepartfamily}{\namepartgiveni}{\namepartprefixi}{\namepartsuffixi}}%
	\or
	\usebibmacro{name:first-last}{\namepartfamily}{\namepartgiven}{\namepartprefix}{\namepartsuffix}%
	\fi
	\usebibmacro{name:andothers}%
	\ifnumequal{\value{listcount}}{\value{liststop}}{'s}{}}
\DeclareFieldFormat{shorthand:poss}{%
	\ifnameundef{labelname}{#1's}{#1}}
\DeclareFieldFormat{citetitle:poss}{\mkbibemph{#1}'s}
\DeclareFieldFormat{label:poss}{#1's}
\newrobustcmd*{\posscitealias}{%
	\AtNextCite{%
		\DeclareNameAlias{labelname}{labelname:poss}%
		\DeclareFieldAlias{shorthand}{shorthand:poss}%
		\DeclareFieldAlias{citetitle}{citetitle:poss}%
		\DeclareFieldAlias{label}{label:poss}}}
\newrobustcmd*{\posscite}{%
	\posscitealias%
	\textcite}
\newrobustcmd*{\Posscite}{\bibsentence\posscite}
\newrobustcmd*{\posscites}{%
	\posscitealias%
	\textcites}

\DeclareFieldFormat*{edition}{\nth{#1} ed.}

\newcommand{\noop}[1]{}

\usepackage{microtype}

\usepackage{siunitx} % centering in tables

\def\yyy{%
	\bgroup\uccode`\~\expandafter`\string-%
	\uppercase{\egroup\edef~{\noexpand\text{\llap{\textendash}\relax}}}%
	\mathcode\expandafter`\string-"8000 }

\def\xxxl#1{%
	\bgroup\uccode`\~\expandafter`\string#1%
	\uppercase{\egroup\edef~{\noexpand\text{\noexpand\llap{\string#1}}}}%
	\mathcode\expandafter`\string#1"8000 }

\def\xxxr#1{%
	\bgroup\uccode`\~\expandafter`\string#1%
	\uppercase{\egroup\edef~{\noexpand\text{\noexpand\rlap{\string#1}}}}%
	\mathcode\expandafter`\string#1"8000 }

\usepackage{setspace}
\usepackage[bottom]{footmisc}
\usepackage[framemethod=tikz]{mdframed}
\usepackage{tikz}
\usetikzlibrary{shapes.geometric, arrows, positioning, calc, decorations.markings, shapes.misc, fit}

\usepackage{acronym}
\usepackage{etoolbox}
\makeatletter
\patchcmd{\maketitle}{\@makefntext}{\fakecommand}{}{}
\patchcmd{\maketitle}{\rlap}{\hbox}{}{}
\patchcmd{\@maketitle}{\@author}{\hspace*{5pt}\@author}{}{}
\makeatother

\usepackage{comment}

\usepackage{changes}

\definechangesauthor[color=Green]{NA}

\usepackage{todonotes}
\usepackage{abstract}
\usepackage{titling}

\usepackage[noblocks]{authblk}

\title{\noindent\justifying Do conditional cash transfers in childhood increase economic resilience in adulthood? Evidence from the COVID-19 pandemic shock in Ecuador\thanks{Corresponding author: Jos{\'e}-Ignacio Ant{\'o}n, Department of Applied Economics, University of Salamanca, Campus Miguel de Unamuno, 37007 Salamanca (Spain), e-mail: \href{mailto:janton@usal.es}{\texttt{janton@usal.es}}.}}

\pretitle{\vspace{-40pt}\centering\Large\bfseries}
	\posttitle{\par\centering}

\predate{\centering}
	\postdate{\centering\vskip -2em}
\makeatletter
\AtBeginDocument{%
	\newlength{\temp@x}%
	\newlength{\temp@y}%
	\newlength{\temp@w}%
	\newlength{\temp@h}%
	\def\my@coords#1#2#3#4{%
		\setlength{\temp@x}{#1}%
		\setlength{\temp@y}{#2}%
		\setlength{\temp@w}{#3}%
		\setlength{\temp@h}{#4}%
		\adjustlengths{}%
		\my@pdfliteral{\strip@pt\temp@x\space\strip@pt\temp@y\space\strip@pt\temp@w\space\strip@pt\temp@h\space re}}%
	\ifpdf
	\typeout{In PDF mode}%
	\def\my@pdfliteral#1{\pdfliteral page{#1}}% I don't know why % this command...
	\def\adjustlengths{}%
	\fi
	\ifxetex
	\def\my@pdfliteral #1{}% isn't equivalent to this one
	\def\adjustlengths{\setlength{\temp@h}{-\temp@h}\addtolength{\temp@y}{1in}\addtolength{\temp@x}{-1in}}%
	\fi%
	\def\Hy@colorlink#1{%
		\begingroup
		\ifHy@ocgcolorlinks
		\def\Hy@ocgcolor{#1}%
		\my@pdfliteral{q}%
		\my@pdfliteral{7 Tr}% Set text mode to clipping-only
		\else
		\HyColor@UseColor#1%
		\fi
	}%
	\def\Hy@endcolorlink{%
		\ifHy@ocgcolorlinks%
		\my@pdfliteral{/OC/OCPrint BDC}%
		\my@coords{0pt}{0pt}{\pdfpagewidth}{\pdfpageheight}%
		\my@pdfliteral{F}% Fill clipping path (the url's text) with
		\my@pdfliteral{EMC/OC/OCView BDC}%
		\begingroup%
		\expandafter\HyColor@UseColor\Hy@ocgcolor%
		\my@coords{0pt}{0pt}{\pdfpagewidth}{\pdfpageheight}%
		\my@pdfliteral{F}% Fill clipping path (the url's text)
		\endgroup%
		\my@pdfliteral{EMC}%
		\my@pdfliteral{0 Tr}% Reset text to normal mode
		\my@pdfliteral{Q}%
		\fi
		\endgroup
	}%
}
\makeatother

\DeclareLabeldate{%
	\field{date}
	\field{eventdate}
	\field{origdate}
	\field{urldate}
	\field{pubstate}
	\literal{nodate}
}

\renewbibmacro*{addendum+pubstate}{%
	\printfield{addendum}%
	\iffieldequalstr{labeldatesource}{pubstate}{}
	{\newunit\newblock\printfield{pubstate}}}

\author[$\dag$]{Jos{\'e}-Ignacio~Ant{\'o}n}
\affil[$\dag$]{University of Salamanca (Spain)}
\author[$\ddag$]{Ruthy~Intriago}
\affil[$\ddag$]{FLACSO-Ecuador (Ecuador)}
\author[$\ddag$]{Juan~Ponce}

\usepackage[pdftex,pdftitle={},pdfsubject={},%
pdfauthor={},breaklinks=true]{hyperref}
\hypersetup{
	colorlinks=true,%
	citecolor=blue,%
	filecolor=black,%
	linkcolor=blue,%
	urlcolor=blue}

\begin{document}

%\date{This version: \today}
\date{}

\maketitle

\singlespacing	

%\vspace*{-1cm}

\begin{abstract}
	\noindent The primary goal of conditional cash transfers (CCTs) is to alleviate short-term poverty while preventing the intergenerational transmission of deprivation by promoting the accumulation of human capital among children. Although a substantial body of research has evaluated the short-run impacts of CCTs, studies on their long-term effects are relatively scarce, and evidence regarding their influence on resilience to future economic shocks is limited. As human capital accumulation is expected to enhance individuals' ability to cope with risk and uncertainty during turbulent periods, we investigate whether receiving a conditional cash transfer---specifically, the Human Development Grant (HDG) in Ecuador---during childhood improves the capacity to respond to unforeseen exogenous economic shocks in adulthood, such as the COVID-19 pandemic. Using a regression discontinuity design (RDD) and leveraging merged administrative data, we do not find an overall effect of the HDG on the target population. Nevertheless, we present evidence that individuals who were eligible for the programme and lived in rural areas (where previous works have found the largest effects in terms of on short-term impact) during their childhood, approximately 12 years before the pandemic, exhibited greater economic resilience to the pandemic. In particular, eligibility increased the likelihood of remaining employed in the formal sector during some of the most challenging phases of the COVID-19 crisis. The likely drivers of these results are the weak conditionality of the HDG and demand factors given the limited ability of the formal economy to absorb labour, even if more educated.
	\vskip 0.5em
	
	\vskip 0.5em
	
\end{abstract}

\FloatBarrier
\section{Introduction}\label{Section 1}

Since Mexico launched \textit{Progresa} in 1997, many countries across five  continents have followed its lead and have introduced CCTs. By 2018, 63 states worldwide were operating a programme of this kind \parencite{wb2018}. In many cases, these benefits are the national flagship social intervention for combatting poverty. The most salient feature of these programmes is that welfare payments are conditional on meeting requirements such as ensuring children attend school and attend medical check-ups. Therefore, their primary objective is to improve economic outcomes for the next generation, thereby preventing the intergenerational transmission of poverty. While a large body of research evaluates the short-term effects of CCTs and reaches optimistic conclusions, particularly regarding educational enrolment \parencite{fiszbein2009,marshall2015,pasha2024}, research on their long-run impact remains scarce and evidence on their impact on resilience to future economic shocks is lacking.

In this paper, we examine whether receiving a conditional cash transfer---the HDG in Ecuador---during childhood enhances individuals' capacity to respond to unanticipated exogenous economic shocks, specifically the COVID-19 pandemic, in adulthood. The rationale behind this question is that the transfer contributes to the accumulation of human capital, which may increase individuals' ability to manage risk and uncertainty \parencite{schultz1961, schultz1975}. Educated workers are better equipped to adapt to changing working conditions and acquire new skills. It is therefore reasonable to expect that individuals with higher levels of human capital could navigate the COVID-19 crisis more effectively---e.g., by adapting to remote working technologies or transitioning to jobs requiring technological skills \parencite{fasih2020}. Furthermore, the higher costs associated with matching skilled workers to jobs, in terms of both hiring and firing, suggest that human capital acts as a buffer against macroeconomic turbulence \parencite{ljungqvist2007}. Extensive empirical evidence supports the plausibility of this argument, both historically \parencite{bell2011, cutler2015, mu2006, leung2014, crescenzi2016} and during the COVID-19 crisis \parencite{adams2020, casarico2022, farre2022, laurimae2024, qian2020}.

In light of the evidence yielded by previous research on this social transfer in Ecuador, this question takes special importance. This literature suggests a significant impact on school enrolment \parencite{schady2008a,oosterbeek2008}, and even on a child's probability of completing secondary education roughly six and ten years after becoming eligible for the programme, respectively \parencite{araujo2018}. Similarly, prior literature also suggests that the grant reduced child labour \parencite{schady2006,martinez2012}. However, it does not provide conclusive findings on the impact on child development or health, which appears to be limited to certain population segments only \parencite{paxson2010,ponce2010,fernald2011}. The grant does not appear to have exerted any impact on test scores, employment rates or labour market outcomes in the formal economy 10--15 years after eligibility nfor the programme, with the notable exception of the performance of the non-mestizo population, mainly composed by indigenous and Afro-Ecuadorians, in terms of their integration into the formal sector \parencite{araujo2018,ponce2025}.

To shed light on this issue we combine information from 2008/2009 poverty census (on which the programme's targetin is based) with social security records from February 2020 to February 2023. Using a RDD that exploits the discontinuity in eligibility around a cut-off of a poverty index included in the aforementioned cadastre, we estimate the local intention-to-treat (ITT) effect of HDG eligibility in 2008/2009 on the probability of maintaining a formal job month by month during the three years following the outbreak of the pandemic in Ecuador in February 2020.

Our results indicate that receiving the programme in childhood had no overall effect on economic resilience during the pandemic in Ecuador. However, we find that the HDG clearly increased the probability of remaining employed in the formal sector during the pandemic for individuals living in rural areas at the time of receiving the CCT. Previous literature has identified precisely these places as having the largest short-term impact of the HDG on both education and health.  Eligible individuals within this population group were more likely to keep a formal job than ineligible ones from roughly July 2020 to September 2021, when the latter group recovered and caught up with the former. These positive long-term effects were driven by male workers.

Our work contributes to the broader literature on the long-term impact of CCT and to the body of research focused on the effects of social programmes on resilience to economic shocks. A number of works have examined the impact of receiving a CCT in childhood on education, fertility and labour market outcomes in adulthood \parencite{araujo2018,araujo2021,alam2011,baez2011,barham2024,behrman2011,garcia2012,kugler2018,laguinge2024,molina2020,parker2023,ponce2025}, with mixed results. Similarly, the amount of research investigating how social programmes, including CCTs, help people to cope with sudden external shocks (like weather-related extreme events, natural disasters or health crisis) is considerable \parencite{asfaw2017,bandiera2019,banerjee2020,bottan2021,canedo2023,carraro2023,christensen2020,dejanvry2006,flaminiano2021,haile2022,karlan2022,londono2022,matata2023,premand2022,rosas2022}. 

Nevertheless, only a few studies have studied the effect of interventions on the long-term ability to deal with these types of adverse events. \textcite{casey2023} explore the effect of a community-driven development programme in Sierra Leone on the response to the 2014 Ebola epidemic 11 years later. The work of \textcite{fiala2025} focuses on the effect of a one-time cash transfer business creation subsidy in Uganda on employment and income during lockdowns imposed to combat the COVID-19 virus. To the best of our knowledge, this is the first study to explore the impact of benefiting from a CCT in childhood on employment resilience during an exogenous economic shock in adulthood. 

Our paper does not only contribute to the mentioned research stream but also adds to the debate raised by \textcite{leight2022}, who warned against the common practice---and potential bias---of rarely conducting long-term follow-ups on social interventions when short-run results are modest, as has sometimes been the case with CCT programmes. This is indeed the case with the HDG, which has had less of an impact on human capital accumulation than other similar programmes in Latin America and the Caribbean, and has had no impact on subsequent employment and income overall \parencite{araujo2018,ponce2025}. 

The rest of the paper unfolds as follows. Section~\ref{Section 2} provides background on the Ecuadorian labour market and the country's experience of COVID-19 crisis. Section~\ref{Section 3} discusses our research design, including the institutional setting, data and empirical strategy. We present the results of our analyses in Section~\ref{Section 4}, while Section~\ref{Section 5} summarises and discusses the main implications of the research.

\FloatBarrier
\section{Background}\label{Section 2}

When evaluating the performance of Ecuadorian workers in the formal labour market, it is helpful to first outline the characteristics of this sector in Ecuador. Th formality definition used here follows the \enquote{legalistic} or \enquote{social protection} approach \parencite{gasparini2009}. Formal workers are those affiliated with social security, who consequently enjoy rights to certain social benefits, like contributory old-age, survivors' and disability pensions, maternity and sickness benefits, unemployment insurance and healthcare \parencite{ssa2020}. Informality is a multidimensional phenomenon associated with various explanatory factors (economic structure, lack of law enforcement, poor public services and burdensome regulatory frameworks). Whether participation in the informal labour market is a voluntary decision is also debated \parencite{loayza2009,maloney2004,cimoli2006,portes1993,biles2009}.

Informality is the main problem in the Ecuadorian labour market, being pervasive and almost endemic. While unemployment represented 3.4\% of the labour force, this segment of the labour market accounted for 63.5\% of total employment in the last quarter of 2019, right before the outbreak of the pandemic \parencite{ilo2025}. Regardless of the reasons for the existence of this sector, informal workers in Ecuador have much worse social outcomes on average than those in the formal economy, such as lower earnings, higher poverty rates and worse future career prospects \parencite{canelas2019,matano2020,maurizio2019,maurizio2021,maurizio2023}. Furthermore, available evidence suggests that informality can result in negative externalities as well. A larger informal sector not only leads to higher inequality but also has a negative impact on tax collection \parencite{boitano2019}, the health status of the population \parencite{utzet2021}, pension coverage \parencite{daude2015} and political participation \parencite{baker2022}. Therefore, working in the formal sector in Ecuador is definitely a positive socio-economic outcome that is of unquestionable interest to policymakers. 

\begin{figure}[p]
	\footnotesize
	\caption{Excess mortality from the COVID-19 in Ecuador}
	\captionsetup{width=0.75\textwidth}
	\label{Figure 1}
	\centering
	\includegraphics[width=0.75\textwidth]{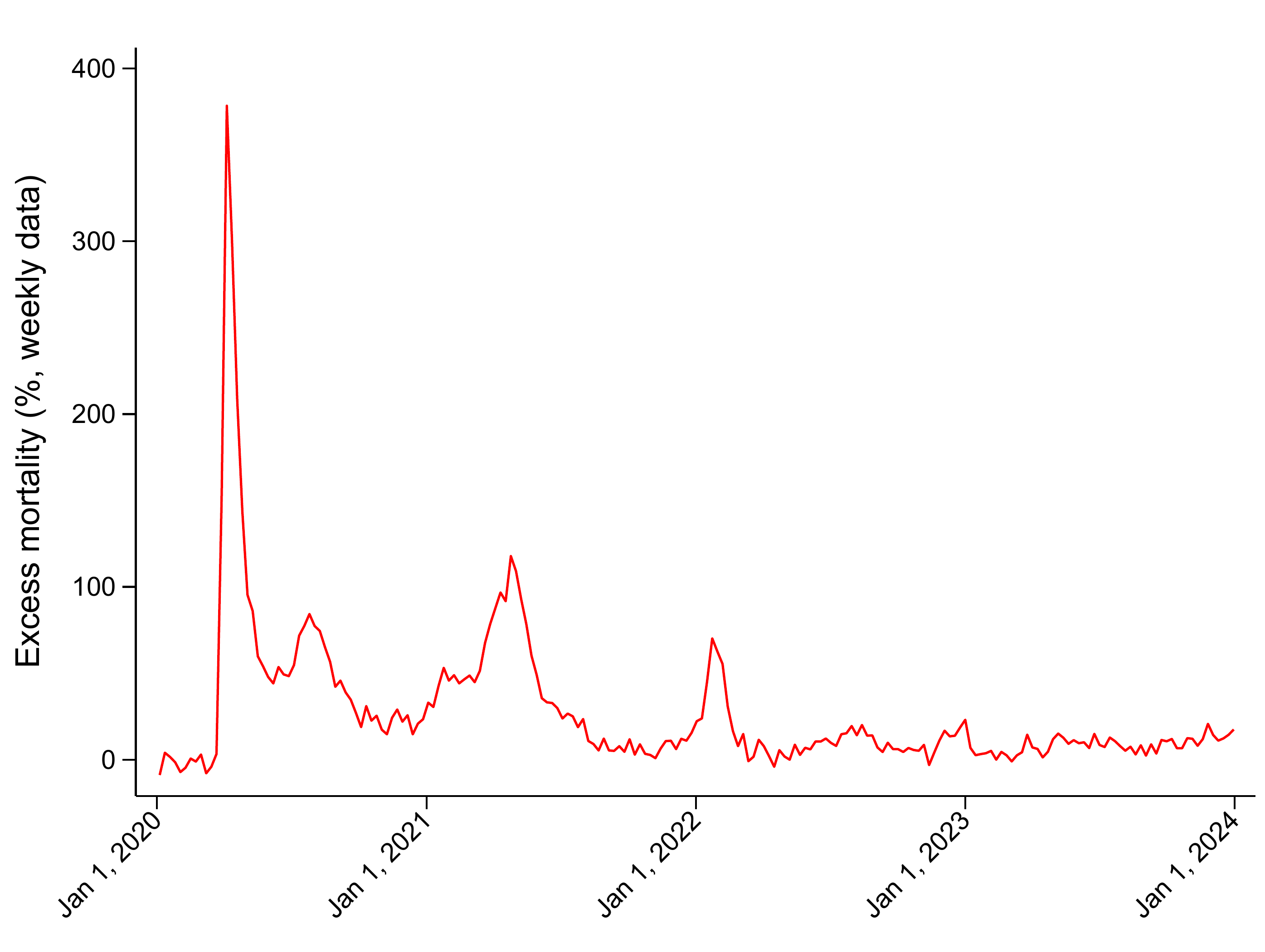} \\
	\noindent\makebox[\textwidth][c]{
		\noindent\begin{minipage}[c]{0.75\textwidth}	
			\justifying
			\noindent\textit{Notes}: The figure shows weekly P-scores, defined as the percentage difference between the reported number of weekly or monthly deaths in 2020-–2023 and the projected number of deaths for the same period based on previous years.
			\noindent\textit{Source}: Authors' analysis from \textcite{owd2025a}.
		\end{minipage}
	}
\end{figure}

The COVID-19 pandemic had a severe impact on Ecuador's health and economy. Out of the 95 countries for which data are available, Ecuador showed the largest cumulative excess mortality (30.7\%) from \nth{1} January 2020 to \nth{5} May 2023 (when the World Health Organization declared the end of the pandemic) \parencite{owd2025b}. This impressive figure is largely due to the lethality of the first wave, which peaked at the beginning of April 2020 (see Figure~\ref{Figure 1}).\footnote{The usefulness of excess of mortality for tracking of the pandemic is related to the differences in testing and death reporting availability over time and across countries \parencite{beaney2020}. This measure also includes non-COVID-19 deaths, such as those from other health conditions that went untreated due to the healthcare system being overwhelmed. Furthermore, Ecuador experienced severe delays in reporting both new cases and deaths. The government later corrected its figures, periodically adding a large number of missing cases and deaths related to the virus, which created artificial peaks in the series and reduced the usefulness of figures for cases and deaths from official sources (we show them in Figure~\ref{Figure A1} in the Appendix).} 

\begin{figure}[p]
	\footnotesize
	\captionsetup{width=0.8\textwidth}
	\caption{Ecuador government response to COVID-19}
	\label{Figure 2}	
	\centering 
	\medskip
	\begin{subfigure}[t]{1\textwidth}
		\centering
		\captionsetup{width=0.8\textwidth}
		\caption{Stringency Index (0--100)}
		\includegraphics[width=0.8\textwidth]{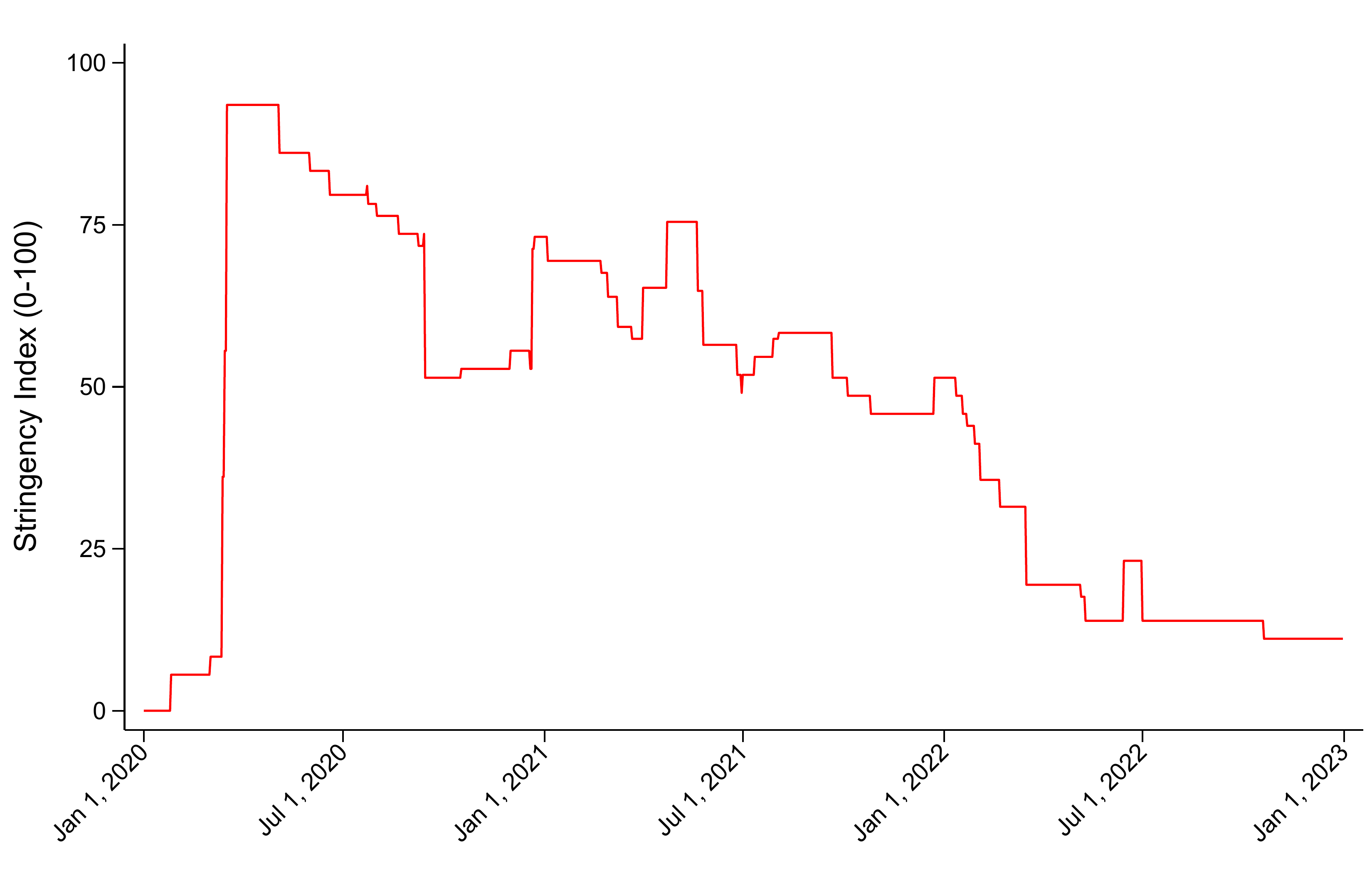} \\
	\end{subfigure}
	\begin{subfigure}[t]{1\textwidth}
		\centering
		\captionsetup{width=0.8\textwidth}
		\caption{Workplace closure policy (0--3)}
		\includegraphics[width=0.8\textwidth]{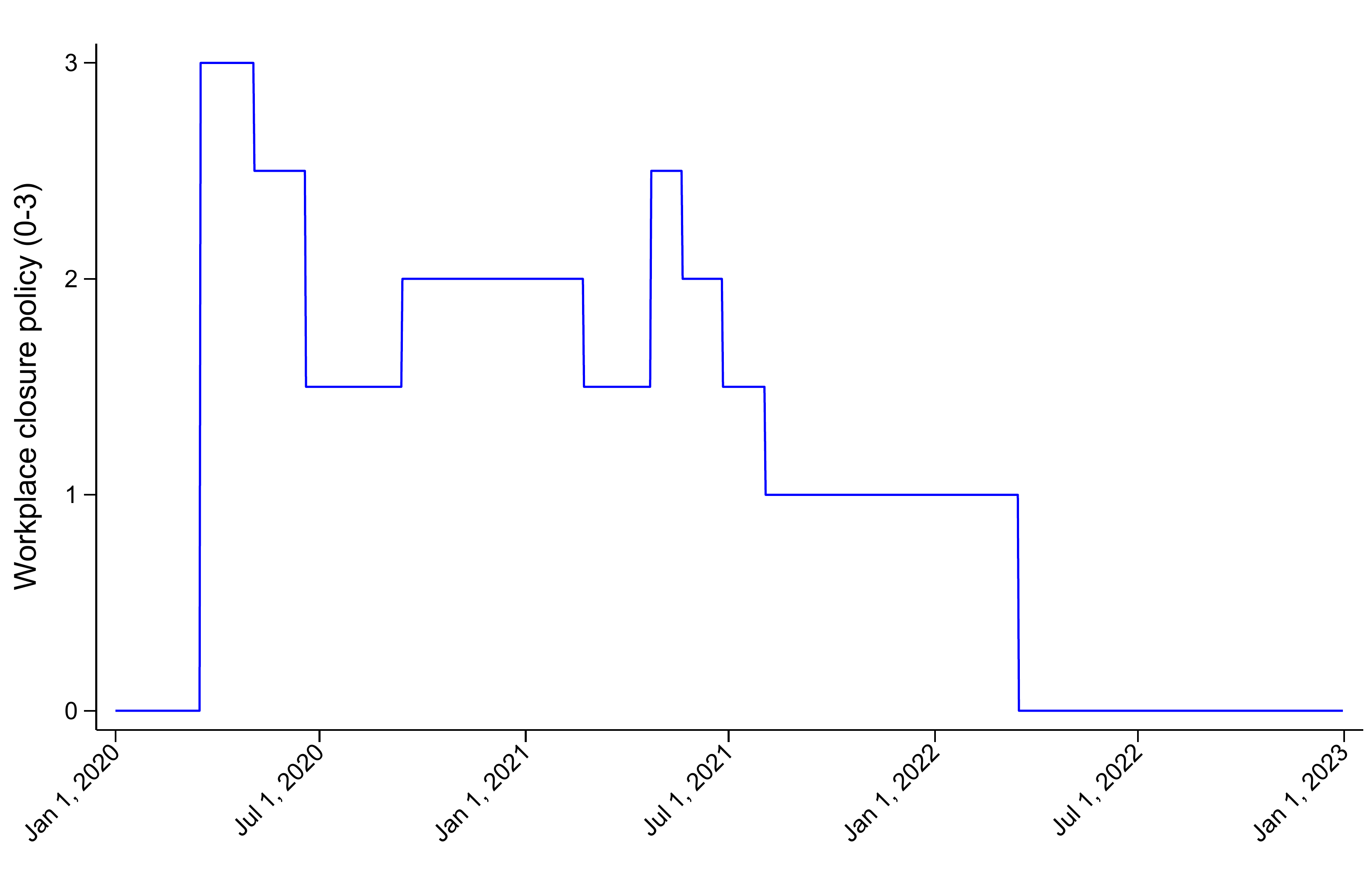} \\
	\end{subfigure}
	\noindent\makebox[\textwidth][c]{
		\noindent\begin{minipage}[c]{0.8\textwidth}	
			\justifying
			\noindent\textit{Notes}: Stringency Index in Panel (a) takes  into account the following metrics: school closures, workplace closures, cancellation of public events, restrictions on public gatherings, closures of public transport, stay-at-home requirements, public information campaigns, restrictions on internal movements and international travel controls. The workplace closure policy depicted in Panel (b) follows the coding scheme below: 0: no measures; 1: recommend closing (or recommend work from home) or all businesses open with alterations resulting in significant differences compared to non-COVID-19 operation; 2: require closing (or work from home) for some sectors or categories of workers; 3: require closing (or work from home) for all-but-essential workplaces (e.g., grocery stores, doctors). Non-integer figures reflects that the next highest policy level only applies to some subnational jurisdictions.\\ 
			\noindent\textit{Source}: Authors' analysis from \textcite{hale2021}.
		\end{minipage}
	}
\end{figure}

Since the beginning of the health crisis in March 2020, the Ecuadorian government issued several presidential decrees to limit mobility and regulate the closure and later (eventual) reopening of activities in the country \parencite{coral2022}. Figure~\ref{Figure 2} shows the overall stringency of the government policies against the virus, particularly those affecting workplaces. These were similar in magnitude to those implemented in other Latin American and Caribbean countries.

The government declared a state of exception nationwide on \nth{16} March. Under the supervision of the Constitutional Court, the executive branch extended that emergency status until \nth{12} September of that year. Hereafter, we refer to the period between \nth{16} March and \nth{12} September 2020---that with the tightest measures in force---as the national lockdown. As a part of a package of policies that included the suspension of constitutional rights (such as freedom of movement and association), a curfew, the closure of shopping malls, restaurants, gyms, cinemas and other leisure facilities, restrictions on vehicle traffic and public transport and the implementation of strict health protocols, the government ordered the cessation of economic activity in non-essential industries: construction, accommodation and food services, education and most of wholesale and retail trade and manufacturing activities. The only sectors not forced to close were those involved in food provision, human health and social work activities, finance and insurance activities and the manufacture of vital goods. Where possible, transitioning to remote learning was implemented, and all in-person educational activities were suspended across the whole country from \nth{12} March 2020 to \nth{14} March 2022. 

Although the government did not enable any financial support for formal firms, it allowed them to make use of various unilateral mechanisms of flexibility: working from home arrangements, reduction of working hours (with proportional wage cuts), changes in working time and suspension of activity (with recovery of lost working hours after the lockdown). To mitigate the impact on vulnerable groups, the government delivered three payments of 60, 60 and {\SI{120}[\$]{}} to roughly \SI{400000} households in poverty not covered by existent programmes between April and May 2020.

The government introduced a \enquote{traffic light system} (red, amber, and green) to adjust restrictions based on local epidemiological conditions. Urban areas with high population density, such as Quito and Guayaquil, were subject to stricter protocols, including intensified control operations and targeted measures in critical sectors. In contrast, rural areas with lower population density were granted greater flexibility, as long as they complied with general prevention guidelines.

The gradual return to normality began in May 2020 with the implementation of a phased economic reactivation plan. The executive branch established specific protocols for each sector, including capacity limits, staggered working hours and biosecurity measures. While economic activity gradually resumed, restrictions on public transport and bans on mass gatherings remained in place.

On \nth{23} April 2021, amid a surge in cases and hospital overcrowding, the presidency declared a new state of exception in 16 provinces. Measures adopted included a nightly curfew and complete weekend lockdowns for one month. However, logistical challenges and public resistance emerged, as evidenced by overcrowding in markets and shopping centres. On \nth{20} May 2021, the central government delegated responsibilities to cantonal committees, enabling them to implement measures based on local epidemiological conditions. This marked the end of the nationwide state of emergency, although the curfew remained in effect under the executive branch's supervision. Finally, on \nth{11} May 2023, the government officially declared the end of the health emergency, incorporating the monitoring of COVID-19 into the routine surveillance of respiratory diseases. 

In 2020, due to the collapse of the external sector and government restrictions on the economic activity, resulted in a 9.3\% contraction in output, one of the most severe in Latin America and the Caribbean \parencite{eclac2025}. Unsurprisingly, the economic turmoil had devastating consequences for the functioning of the national labour market (Figure~\ref{Figure 3}). Informal workers were among the worst affected at the start of the pandemic. Unlike in conventional economic crises, where the informal sector often serves as an exhaust valve for job loss in the formal economy, the prohibition of non-essential activities and the high proportion of workers in contact-intensive industries and the low share of teleworkable jobs in this sector \parencite{imf2020} led to a five-point decline in informal employment by the end of the first (and toughest) closure.
 
Conversely, as in other countries, the formal sector initially served as a refuge \parencite{acevedo2021} due to the flexibility mechanisms enabled by the Ecuadorian government (such as reduced working hours and remote work) and the existence of severance payments which made dismissals costly \parencite{wb2021}. The impact of the shock was uneven across population groups, exacerbating pre-existing labour market inequalities. Overall, during the first months of the health crisis, women, young people, older individuals, less educated groups, people of African descent or indigenous background and migrants were hit hardest by the economic and labour market turmoil \parencite{eclac2021b,imf2021,olivieri2023}.

Informal employment quickly rose and surpassed pre-pandemic levels as soon as restrictions on economic activity began to ease. However, employment in the formal sector plummeted after the state of exception ended in September 2020, due to the removal of flexible working arrangements and the closure of firms that were unable to recover following the reopening of the economy. Although GDP and total employment surpassed their 2019 levels by the end of 2021, the crisis left a lasting impact on the formal sector, with employment levels remaining 10\% lower than in February 2020. By the end of 2024, formal employment remained below pre-pandemic levels. This relatively exceptional dynamic in the region appears to be due to the severity of the contraction in manufacturing and construction, which absorb most of the formal employment in the country
    
\FloatBarrier   
\section{Research design}\label{Section 3}

\FloatBarrier
\subsection{Institutional setting}\label{Subsection 3.1}

The HDG started in Ecuador in 2003 as a reformulation of the Solidarity Grant, a poorly targeted earlier social benefit introduced in 1998 as a safety net for socially disadvantaged families. This grant aimed to compensate socially disadvantaged these households for the removal of subsidies on gas, petrol and electricity, which was part of the liberalisation and adjustment policies adopted by the country in the late 1990s. The HDG programme aspires to alleviate poverty in the short term and foster human capital formation in order to prevent the intergenerational transmission of poverty. Specifically, it targets vulnerable families with children under the age of 16. 

The redesign of the programme in 2003 was aimed at improving its targeting. In particular, the Ecuadorian government---with technical assistance from universities---created an ad hoc poverty census (in Spanish, \textit{Sistema de Identificaci{\'o}n y Selecci{\'o}n de Beneficiarios de Programas Sociales} [SELBEN]) to identify the most vulnerable population more accurately. The data collection on households' socio-economic and demographic characteristics relied on visits, public calls and the option for households to sign up to the registry on a voluntary basis and to request that the government evaluate their eligibility for inclusion. 

Initially, the national authorities initially developed an eligibility index ranging from 0 to 100 (from the lowest to the highest level of well-being) based on a principal component analysis of 27 household variables. Households in the first two quintiles of the index with children under 16 years old were eligible for the grant.

In principle, the programme imposes two conditions on grant beneficiaries. The first requirement relates to education: children aged six to 18 must be enrolled in school and maintain monthly attendance of at least 80\%. The second condition relates to healthcare: infants and children up to five years old must attend a series of medical check-ups (one preventive check-up every two months for children under one year old and one every six months for children between one and five years old). Unlike most of the other CCTs operating in Latin America and the Caribbean, this programme has set up no enforcement mechanism to verify that beneficiaries meet these conditions. Consequently, the Ecuadorian authorities do not suspend the benefits if families fail to comply with the requirements, meaning that the transfer is effectively unconditional in practice. While the Ecuadorian government does not verify compliance, families do commit in writing to satisfying the conditions, and the authorities have always publicly emphasised the importance of meeting the these requirements.

In theory, disregarding the non-enforcement of conditionality, loss of eligibility due to the absence of children under the required age or a poverty index above the cut-off point implied withdrawal of benefits. In practice, government employees regularly visited households to update the poverty census. This process could result in the HDG being suspended. Unfortunately, information on how often this occurred is scarce. For example, the government maintained the grant for households with children under 18. Loss of eligibility under this criterion should be easy to monitor, but updates to eligibility status can take months or years to be implemented in practice. Furthermore, Ecuadorian authorities tend to withdraw grants from large numbers of households at once rather than continuously on a case-by-case basis.    

At the programme's outset, the government committed to renewing the registry approximately every five years and, to this end, set up the Social Registry 2008/2009. Administered by the Social Registry Unit, a public agency under the auspices of Ecuador's Ministry of Economic and Social Inclusion, the Social Registry operated similarly to SELBEN and aimed to improve the HDG's targeting. Since the 2008/2009 wave, the National Institute of Statistics and Censuses of Ecuador has been responsible for data collection. At that time, the Social Registry Unit developed a Social Registry Index using non-linear component analysis and 30 household variables. This index was rescaled from 0 to 100 to determine eligibility for different social benefits. The government set the cut-off point at 36.5987. Along with the rules relating to the presence of children under 16, this criterion determined eligibility for the grant from August 2009 to August 2014, when the new Social Registry 2013/14 came into force. Updating the database meant that HDG payments ceased for over 200,000 households \parencite{buser2017}.

The HDG has not only been the flagship social programme of successive Ecuadorian governments for more than two decades, but it has also become one of the most significant initiatives of its kind in Latin America and the Caribbean. In 2022, it represented nearly 1\% of the GDP and reached 7.6\% of the population (more than 12\% in 2011). From 2003 to 2007, the HDG amount was set at \SI{15}[US\$]{} per month, increasing to \SI{30}[US\$]{} in 2008 and then to \SI{35}[US\$]{} between 2009 and 2011. Since 2012, it has remained at \SI{50}[US\$]{}. Starting in 2018, it varies according to the number of children, reaching up to \SI{150}[US\$]{} \parencite{eclac2025}. Bearing in mind that the average national per capita income and the average per capita income of the first two quintiles in 2008 (the year when our analysis starts) were {\SI{166}[\$]{}} and {\SI{32}[\$]{}} per month, respectively \parencite{cedlas2025}, the size of the transfer is considerable. Whenever possible, the recipients are the mothers, who can withdraw the benefit from private banks. In a recent change, eligible mothers can now receive the payment directly into their bank account.  

\FloatBarrier
\subsection{Data}\label{Subsection 3.2}

Our analysis leverages data from two different sources. The first is the Social Registry 2008/2009 database \parencite{socialregistry2025}. As explained above, this cadastre, which is administered by the Ministry of Economic and Social Inclusion, contains static socio-economic and demographic information of Ecuadorian households. It enables public institutions to determine the eligibility of social benefit recipients and ensure appropriate targeting based on the aforementioned poverty index.

The Social Registry database includes the Social Registry Index (and the component variables needed for its calculation) that governs the receipt of the HDG. From August 2009 onwards, households were eligible if they scored fewer than 36.5987 points on the index. The Social Registry 2008/2009 index---hereafter referred to as the poverty index---is therefore the forcing or running variable in our analysis. Furthermore, we have access to information on the actual receipt of the HDG in 2010 according to administrative sources.

We merge these data from the Social Registry with information from our second data source: the \textcite{socialsecurity2025} from February 2020 to February 2023. This source provides information on individuals' registration status on a monthly basis. Consequently, this administrative registry enable us to track the formal sector employment of all individuals included in the Social Registry in 2008/2009 who had a valid identity card number.

Using these data, we can determine the HDG eligibility status of children aged 15 years and under at the time of the survey in 2008/2009, as well as their subsequent participation in the formal sector during the pandemic. 4.8\% of individuals were interviewed in 2007, so our sample comprises children who were at least seven years old at the time of the survey. 78.1\% and 17.1\% of individuals were interviewed in 2008 and 2009, respectively. From now on, the age group configuration aims to produce intervals of a similar size. Our analysis focuses on individuals living with families listed in the 2008/2009 Social Registry whose poverty index was between 30.5 and 42.5 points, who were employed in the formal sector in February 2020 (the start of the pandemic), and who were at least 20 years old at that time. We observe their participation in the formal labour market for three years, until February 2023, to see who remains employed in the formal economy during this period. We focus on this group, leaving aside those not in the labour force, because individuals of this age may still be in education, which is a more positive outcome than employment itself.

The two databases were merged based on the identity card numbers of the children. We assume that individuals with a valid identity card number who are not found in the Social Security records did not participate in formal labour activities. Unfortunately, a significant proportion of individuals (21.3\%) have missing identity card numbers. This may be because it was not collected by the cadastre administrators or because a significant proportion of children did not have a personal identity card in 2008 (as these were not compulsory for children at that time).

Furthermore, the proportion of children without a personal identity card number is significantly higher above the threshold (28\%) than below it (15.8\%). This is due to two reasons. Firstly, surveyors placed greater emphasis on collecting detailed information from individuals below the threshold (i.e. the potential recipients of the benefit) than from those above it. The latter group received much less attention from the Ecuadorian authorities, as they were not relevant to the administration of the programme. Secondly, programme administrators encouraged members of households receiving the benefit to obtain a personal identity card if they did not have one.

In principle, if the probability of not having a valid identity card number were independent of future labour market outcomes, the presence of missing values would not be a major problem and complete case analysis would be a reasonable option. However, the much higher incidence of not having a valid personal identity number above the threshold introduces a discontinuity in the poverty index density at the cut-off. We comment further on this issue in the next section.

To overcome this problem, we impute the participation in the formal labour market between February 2020 and February 2023 using the hot deck method. In short, this technique involves defining strata based on a set of categorical variables, for each respondent with missing values for the variable(s) of interest (the recipient), finding a respondent (the donor) with complete values within the same stratum and taking its values for the variables with missing values \parencite{andridge2010,yan2008}. Advantages of the hot deck method include efficiency gains compared to the complete case analysis, reduced sensitivity to model misspecification compared parametric approaches and that only plausible values can be imputed. We implement a random version of the hot deck method, where missing data are imputed stochastically using the approximate Bayesian bootstrap method proposed by \textcite{rubin1986}.

Our imputation makes use of a set of variables comprising covariates that are likely to affect the outcome (employment in each month) or that are potentially associated with non-response. Specifically, we consider the following variables: position in relation to the cut-off point, gender, age, ethnicity (mestizos and non-mestizos), education level of the household head, marital status of the household head, household size, area of residence and province. To define the strata, we include all of these variables as binary categorical ones.\footnote{In particular, the cells are defined by the following categories: below vs. above the cut-off, males vs. females, 7--11 years old vs. 12--15 years old, mestizos vs. non-mestizos, household heads with up to six years of education vs. more than six years of schooling, single vs. married or in union household heads marital status, household size of less than five vs. five or more people, urban vs. rural area of residence and province dummies.} This technique produces unbiased estimates if the missing data are either missing completely at random or missing at random (i.e., the probability that of a line being missing varies only with respect to the categorical variables used for imputation). Using this method, we are able to impute virtually 100\% of the observations---all but 18 out of \num{82671}---with a missing identity card number. Our subsequent calculations are based on this sample with imputed labour market outcomes. 

Moreover, we also present the results of our main analyses based on the complete cases (i.e., based on the children with an identity card number) and using a multivariate imputation procedure to check the robustness of our findings. The latter approach involves using sequential equations to predict the probability of formal employment in a given month from 2020 onwards. This is based on the same set of variables (plus age squared) as in the hot deck procedure, as well as the employment status in all previous months (starting in February 2020). The assumptions for obtaining unbiased estimates are the same as in the hot deck method.\footnote{For instance, the probability of employment in April 2020 depends on the aforementioned set of covariates and employment status in February and March 2020. In this procedure, it is not necessary to use binary categorical variables and we can include directly continuous variables, as well as canton dummies instead of province dummies---which we cannot use in the hot deck method because the size of the cells when using cantons would be too small.}  

Table~\ref{Table 1} displays the descriptive statistics of the labour market outcomes of interest in several selected months between February 2020 and February 2023, as well as the running variable, the treatment variable and the covariates considered in the analysis. Our sample contains \num{80152} individuals that meet the above conditions. 

\begin{figure}[p]
	\footnotesize
	\centering
	\captionsetup{width=0.80\textwidth}
	\caption{Evolution of employment in Ecuador (December 2019 = 100)}
	\label{Figure 3}	
	\centering 
	\includegraphics[width=0.80\textwidth]{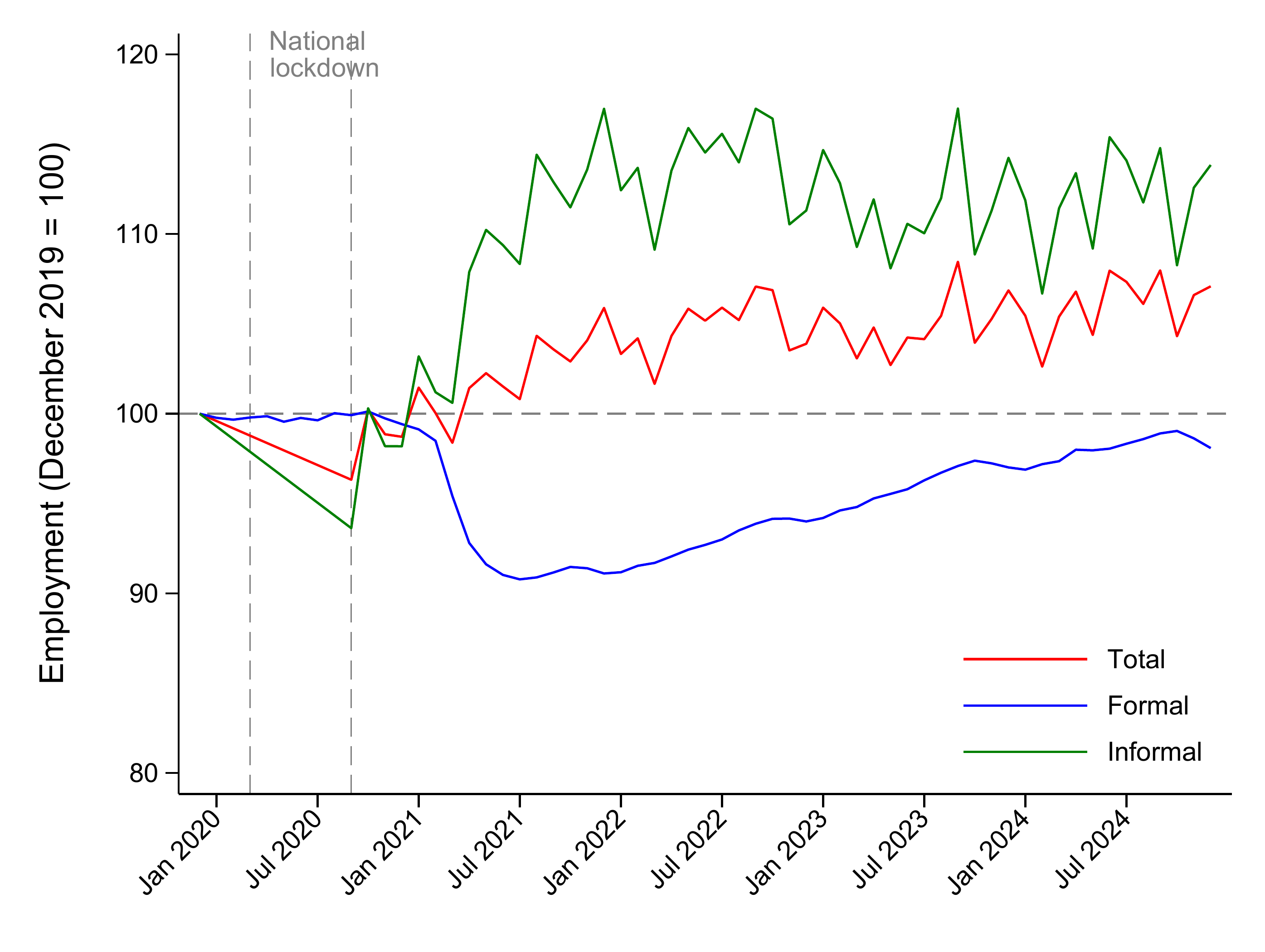}\\
	\noindent\makebox[\textwidth][c]{
		\noindent\begin{minipage}[c]{0.80\textwidth}	
			\justifying
			\noindent\textit{Source}: Authors' analysis from \textcite{inec2025b, inec2025c}.
		\end{minipage}
	}
\end{figure}

\begin{figure}[p]
	\footnotesize
	\centering
	\begin{minipage}[c]{0.80\textwidth}	
		\caption{Proportion of individuals in the sample that remain employed in the formal sector by month}
		\label{Figure 4}
	\end{minipage}
	\centering 
	\includegraphics[width=0.80\textwidth]{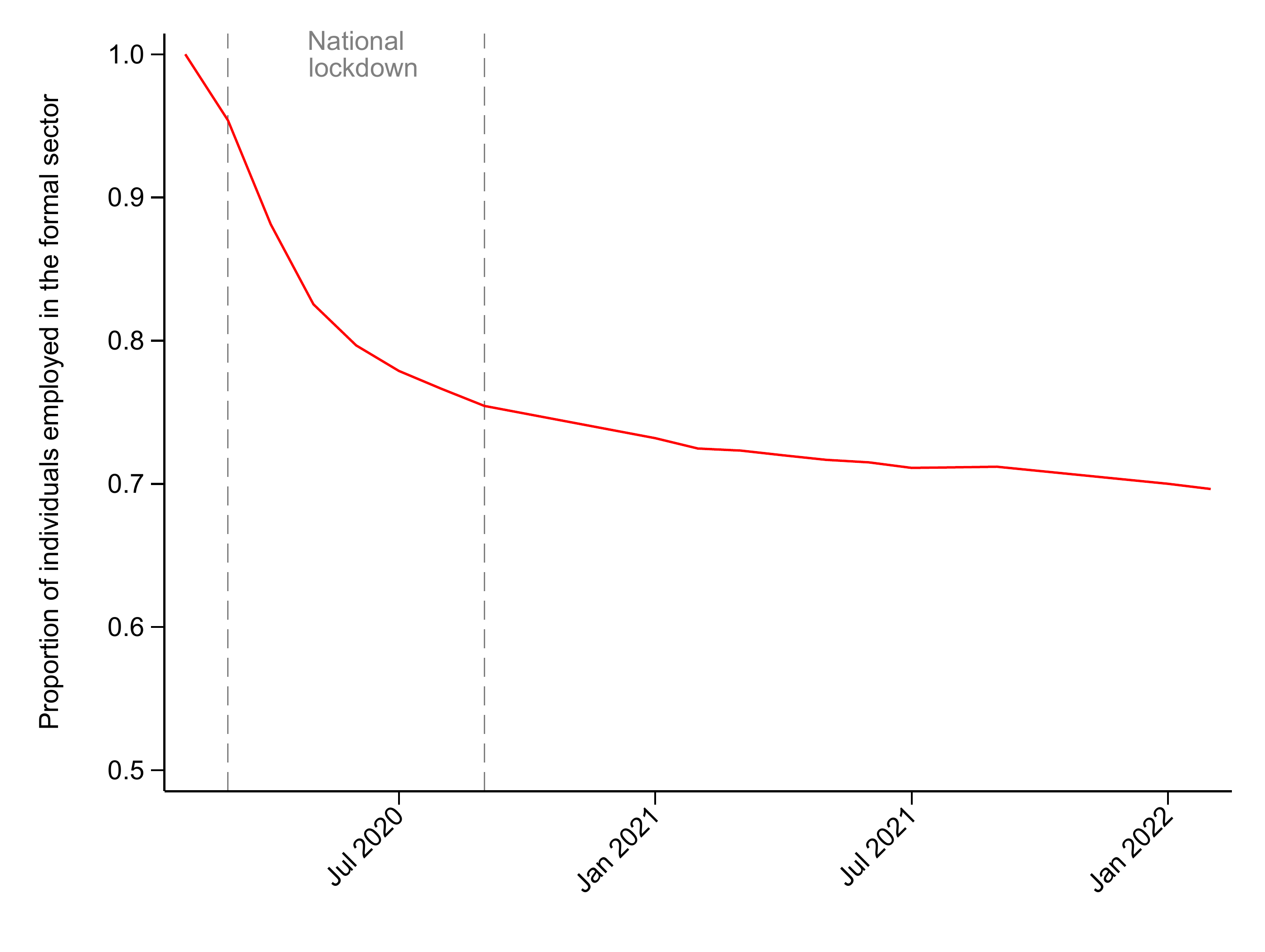}\\
	\noindent\makebox[\textwidth][c]{
		\noindent\begin{minipage}[c]{0.80\textwidth}	
			\justifying\noindent\textit{Source}: Authors' analysis from \textcite{socialregistry2025} and \textcite{socialsecurity2025}.
		\end{minipage}
	}
\end{figure}

Figure~\ref{Figure 4} depicts the proportion of persons in our sample that were employed in February 2020 and remained employed in each subsequent month. We can observe that the decline in the share of employed individuals was particularly marked during the first half of the national lockdown, slowing down considerably afterwards. By February 2023, the probability of keeping employment in the formal sector of the persons in our sample was 30\% lower than before the outbreak of the pandemic.\footnote{Note that this pattern is not necessarily contradictory to the overall trends shown in Figure~\ref{Figure 3} for two reasons. Firstly, Figure~\ref{Figure 3} is also the result of inflows not included in Figure~\ref{Figure 4}: persons who were outside the formal sector in February 2020 but who joined it afterwards. Secondly, our sample consists of individuals who were between the second and the third quintile of the poverty index in 2008/2009. This group was much more likely to have been hit by the economic turmoil---which disproportionally hurt more poor families and young people \parencite{imf2021,eclac2021b}---and hence have experienced a worse-than-average outcome. For instance, it is possible that this group had fewer opportunities to work remotely or that they held more contact-intensive jobs than for formal employees from more advantaged socio-economic backgrounds.}   

\FloatBarrier
\subsection{Empirical strategy}\label{Subsection 3.3}

We use a sharp RDD to evaluate the impact of eligibility for the HDG in 2008/2009 on the probability of having remained employed, month by month from March 2020 to February 2023, among individuals employed in the formal sector in February 2020. Our focus is therefore on the intention-to-treat (ITT) effect, using a research design that exploits the discontinuity in eligibility at the cut-off point in the 2008/2009 Social Registry index (36.5987). Most of our analyses rely on non-parametric local polynomial estimation methods using a data-driven mean squared error (MSE) optimal bandwidth \parencite{calonico2019a,calonico2019b,cattaneo2020a,cattaneo2022}. Our rationale for choosing this method is that global high-order polynomials (i.e., a parametric approach) can lead to noisy estimates and poor coverage of confidence intervals and are sensitive to the polynomial degree \parencite{gelman2019}. 

Our application of this method unfolds as follows \parencite{cattaneo2020a}. First, we select a polynomial of order $p$ and a kernel $K(\cdot)$. The standard practice is to choose $p = 1$ (local linear regression). We also employ $p = 2$ (local quadratic regression) as a robustness check. Second, we choose a bandwidth. For our baseline analysis, we select a bandwidth $h$ that minimises the asymptotic MSE of the point estimator \parencite{calonico2019b}. In our robustness checks, we also consider an alternative optimal bandwidth that minimises the coverage error rate (CER) of the robust bias-corrected confidence interval. 

Third, we choose a kernel function to weigh the observations within the relevant interval around the cut-off point ($c$). We opt for a triangular kernel, $K(u) = (1 - \lvert u \rvert)\mathbb{1}(\lvert u \rvert \leq 1)$, which assigns non-negative weights to each transformed observation (centred around the cut-off point and then divided by the selected bandwidth) based on the distance between the observation's score and the cut-off point. This kernel function is MSE-optimal. 

Fourth, for each month from March 2020 to February 2023, we separately estimate a weighted least squares regression, with the weights given by the kernel function, $K\left(\frac{X_i-c}{h}\right)$, of the outcome variable of individual $i$ at time $t$ ($Y_{it}$)---an employment indicator taking the value of one if the person remains employed and zero otherwise---on an intercept, the treatment indicator $T_i$ (which takes the value one if the poverty index in 2008/2009 is equal to or lower than the cut-off point and zero otherwise), the polynomial on the re-centred forcing variable ($X_i - c$), the polynomial on the re-centred running variable interacted with the treatment indicator and the control covariates of interest ($Z_i$), all of them at their 2008/2009 values. 

We formally estimate the parameter capturing the ITT effect ($\hat{\tau}$) through the lens of the following equation:
\begin{equation}
	\label{Equation 1}
	\hat{Y}_{it} = \hat{\alpha} + \hat{\tau}T_i + \hat{\mu}_{+} (X_i - c) + \hat{\mu}_{-} (X_i - c) T_i + Z_i' \hat{\Theta}
\end{equation}
We assess the precision of this point estimate using robust confidence intervals (more conservative than conventional ones) \parencite{calonico2014b}. Within this framework, the inclusion of controls primarily aims to improve the efficiency of the estimates, rather than correcting unexpected imbalances (which would necessitate invoking parametric assumptions on the regression functions to enable extrapolation or redefining the parameter of interest) \parencite{cattaneo2020a}.

Note that if a household was below the relevant threshold, this implies not only that it must have received the grant when the new targeting rule came into force, but also that it was also very likely to have been paid for several months (or even years) until its eligibility status changed (and this was noticed by the Ecuadorian authorities), or until the Social Registry Unit updated the cadastre. Therefore, if a household was below the cut-off implies that it may have received the HDG for several years. According to the administrative registry of payments for 2010 (the only year for which we have access to this type of information), the targeting of the grant (based on database information, for households with an index value between 30.5 and 42.5 points) was excellent, with very low under-coverage (2.6\% of households below the threshold did not receive the grant in 2010) and leakage rates (only 9.6\% of families above the cut-off point received payments).\footnote{Obviously, our figures refer only to the households in the Social Registry. Considering the entire population that is theoretically eligible for the grant, although the take-up rates of the programme are high (65\% in the two poorest quintiles in 2013), travel costs, personal identity stigma and dissatisfaction with the government pose significant barriers to claiming the HDG \parencite{rinehart2017}.} Therefore, our estimate not only captures the (intrinsically interesting) ITT effect, but also provides a reasonable lower bound on the local average treatment effect. 

Furthermore, we estimate the impact of receiving the grant in 2010 using a fuzzy discontinuity design strategy, where we use the discontinuity in take-up rates around the cut-off in the Social Registry Index 2008/2009 to instrument the payment. This approach has limitations: the discontinuity in the cut-off is very likely to be correlated not only with receipt of the HDG in 2010, but also in subsequent years (for which we lack access to the relevant data). Consequently, we do not consider this strategy as capable of offering a better understanding of the impact of the grants. Conversely, this approach is not superior to our exercise of estimating the ITT effect, so we present it merely as a reassuring robustness check, which yields numerically identical effects. 

\begin{figure}[p]
	\footnotesize
	\centering
	\captionsetup{width=0.80\textwidth}
	\caption{Test for manipulation of the assignment variable based on density discontinuity}
	\label{Figure 5}
	\centering 
	\includegraphics[width=0.80\textwidth]{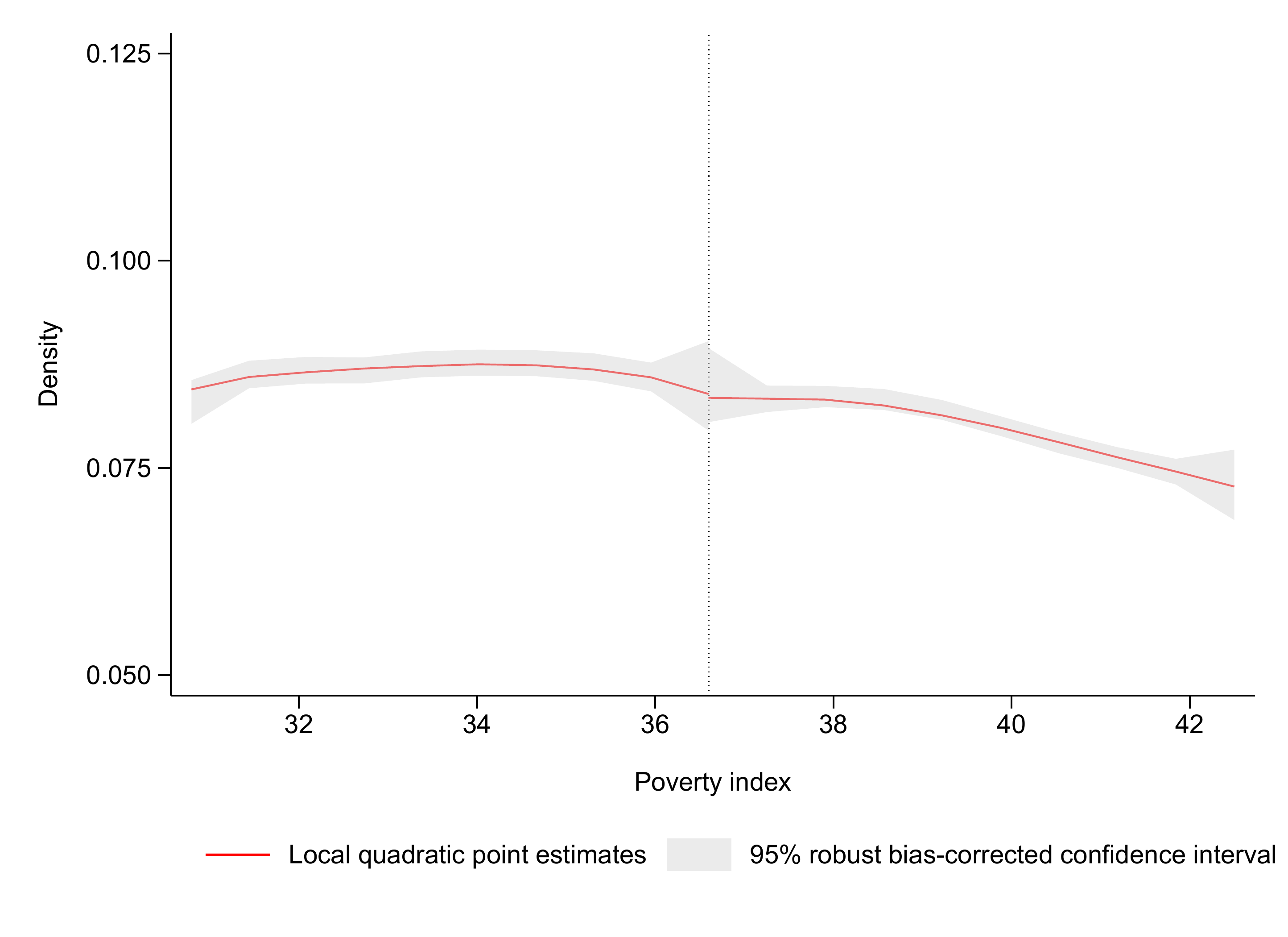}\\
	\noindent\makebox[\textwidth][c]{
		\noindent\begin{minipage}[c]{0.80\textwidth}	
			\justifying
			\noindent\textit{Notes}: The plot explores whether there is a discontinuity in the density of the poverty index using local quadratic regressions estimated separately on either side of the cut-off using a triangular kernel and data-driven MSE-optimal bandwidths. The results of the test do not allow us to reject the hypothesis of continuity ($p-\text{value} = 0.980$).\\   
			\noindent\textit{Source}: Authors' analysis from \textcite{socialregistry2025} and \textcite{socialsecurity2025}.
		\end{minipage}
	}
\end{figure}

The source of identification of the ITT effect is the reduction in households' likelihood of receiving the HDG upon crossing the eligibility threshold in the 2008/2009 Social Registry Index. In other words, the policy application is as good as randomised in the neighbourhood of the cut-off if the research design satisfies certain conditions. The first of these is that there must be no manipulation of the forcing variable by families. Altering their own position relative to the threshold is virtually impossible for families since the Ecuadorian authorities determined the value of the threshold only some time after constructing the entire database. In any case, we perform a manipulation test of the density discontinuity based on local polynomial density methods proposed by \textcite{cattaneo2018,cattaneo2020b}.The results of this test do not allow us to reject the null hypothesis of no manipulation (Figure~\ref{Figure 5}).

The second condition states that there must be no correlation between an observation being below the cut-off point and the factors affecting the labour market outcome. Using the specification outlined above, we assess whether there is any discontinuity in the average values of the observable covariates (measured more than a decade prior to the realisation of the outcome). We find no evidence of discontinuities in these predetermined characteristics at the relevant threshold of the index (Table~\ref{Table 2}). Therefore, we have no reason to anticipate any discontinuity in the corresponding unobservable factors at the cut-off. We repeat the same battery of tests, this time including only individuals with valid identity card numbers, and find that we cannot reject the null hypothesis of continuity at the 10\% level for any of the considered covariates. This suggests that selection into the sample of people with valid identity card numbers based on unobservables is unlikely to be relevant and that our imputation procedures can perform well.

With regard to the complete case sample, we observe a clear jump in the likelihood of having an absent identity card number, as well as a discontinuity in the probability density at the cut-off point (see Figure~\ref{Figure A2}). However, it should be noted that this latter result can hardly be interpreted as evidence of manipulation. In our setting, manipulation is virtually impossible by design. First, the authorities collected all information on household characteristics through interviews. Secondly, they constructed a poverty index using a non-linear principal component analysis method. Finally, they chose as the threshold the figure that left the 40\% poorest households to its left. Thus, the information on households was collected prior to the determination of the cut-off.

A threat to identification in almost any longitudinal study of the effects of a policy on labour market outcomes is selective migration. Unfortunately, we cannot observe this phenomenon in our database, but we can offer some insights into its relevance in our setup. 

Firstly, since we cannot reject continuity in observables, it is reasonable to assume that there is also continuity in the (unobservable) geographical mobility variable. Second, since we use nationwide Social Security records, we are able to control for internal migration, even if we cannot determine who migrated or who did not. Although we cannot identify those who moved, employment spells are recorded as long as individuals are affiliated with the national system.

Thirdly, if the HDG were to affect international migration, then our RDD estimates would be biased. The direction of this bias would depend on whether the impact on migration was positive or negative. For example, if the grant positively impacted migration, we would underestimate the HDG's impact, as international migrants cannot be linked to Social Security records and are generally assumed not to work in the formal economy. Conversely, if the transfer deterred population movements abroad, the RDD estimates would be upward biased. Although we do not have evidence for Ecuador, recent studies suggest that CCTs can foster geographical mobility \mbox{\parencite{araujo2021, parker2023, barham2024}}, although a consensus is lacking \mbox{\parencite[see, e.g.,][]{angelucci2015}}.

Fourthly, although we cannot disentangle this issue from our data, given the undoubted relevance of migration in Ecuador, it is worth discussing this point even further. According to \textcite{olivie2008}, Ecuadorian migrants (who send remittances, i.e., those for whom information is available) have a higher-than-average level of education and come from the richest regions of the country. The HDG targeted the first two quintiles of the poverty index, i.e., the cut-off separated the 40\% poorest from the rest of the population. In principle, one would expect migration to be particularly relevant for socio-economic levels above that marked by the cut-off. Provided we use non-parametric local (not global) polynomial estimation methods, migration should not be a major concern for our analysis.

\FloatBarrier
\section{Results}\label{Section 4}

\FloatBarrier
\subsection{Main results}\label{Subsection 4.1}

In this subsection, we present the main results of our analysis of the impact of being below the index cut-off in 2008/2009 on the probability of having remained employed afte ther outbreak of the COVID-19 among individuals employed in February 2020 (i.e., the ITT effect). We provide both a graphical illustration of the eventual scope of the discontinuity and the econometric results.

\begin{landscape}
	\begin{figure}[p]
		\scriptsize
		\captionsetup{width=0.75\linewidth}
		\caption{Graphical illustration of the RDD: Evaluation of the discontinuity in the proportion of individuals that remained employed in the formal sector in selected months}
		\label{Figure 6}
		\centering 
		\includegraphics[trim={0 0 0 0},width=0.75\linewidth]{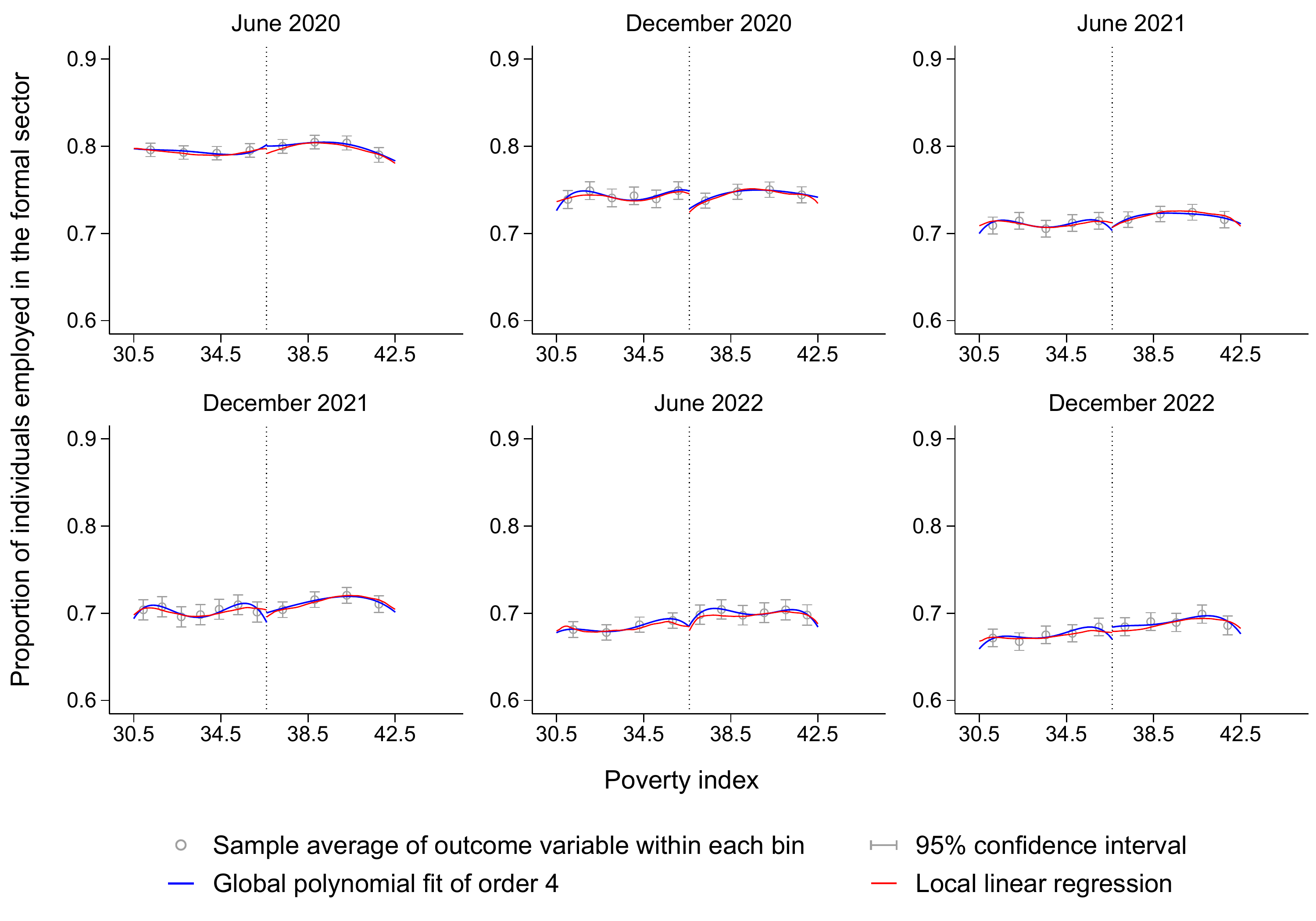}\\
		\noindent\makebox[\textwidth][c]{
			\noindent\begin{minipage}[c]{0.75\linewidth}	
				\justifying
				\noindent\textit{Notes}: The figure shows (i) the sample proportion of individuals employed in the formal sector within each bin and its 95\% confidence interval (grey circles), (ii) the fits of global fourth-order polynomial regressions estimated separately on either side of the cut-off (blue line) and (iii) the fits of local linear regressions estimated separately on either side of the cut-off using a triangular kernel and data-driven MSE-optimal bandwidths (red line). The construction of all plots uses the IMSE-optimal number of disjoint bins.\\ 
				\noindent\textit{Source}: Authors' analysis from \textcite{socialregistry2025} and \textcite{socialsecurity2025}.
			\end{minipage}
		}
	\end{figure}
\end{landscape}

First, as is customary, we graphically assess whether there is a discontinuity in the proportion of persons who remained employed in six selected months after the lockdown started (June 2020, December 2020, June 2021, December 2021, June 2022 and December 2022) (Figure~\ref{Figure 6}). Each subgraph in the figure includes the sample proportion of individuals employed in the formal sector within each bin (with the number of bins determined according to the optimal IMSE criterion), along with its 95\% confidence interval, the fits of global fourth-order polynomial regressions estimated separately on either side of the cut-off and the fits of local linear regressions estimated separately on either side of the cut-off with a triangular kernel and data-driven MSE-optimal bandwidths. 

All three approaches yield similar results: they suggest the absence of any sharp discontinuity in the share of individuals that were able to stay employed in each month included in the analysis. In December 2020, this proportion appears to be slightly higher for those below the cut-off (eligible for the HDG in 2008/2009) than for those the threshold (ineligible for the grant). Consequently, we need to resort to a formal econometric test to determine whether this difference is statistically different from zero. As discussed above, we favour the method based on local linear regressions for the remainder of our analysis, due to its advantages it has over the former, as highlighted in the literature.

We show the estimation results for Equation~\ref{Equation 1} in Table~\ref{Table 3} for the same months selected for the graphical analysis. Aiming at improving the efficiency of our estimates, all the regressions include canton fixed effects and a set of household observable characteristics from 2008/2009. In no month can we reject the null hypothesis that the ITT effect is zero. In the case of December 2020, the estimated coefficient is positive but it is not significant at conventional levels. 

\begin{figure}[p]
	\footnotesize
	\captionsetup{width=0.80\textwidth}
	\caption{ITT effects on the probability of remaining employed in the formal sector month by month}
	\label{Figure 7}
	\centering 
	\includegraphics[trim={0 1.5cm 0 0}, width=0.80\textwidth]{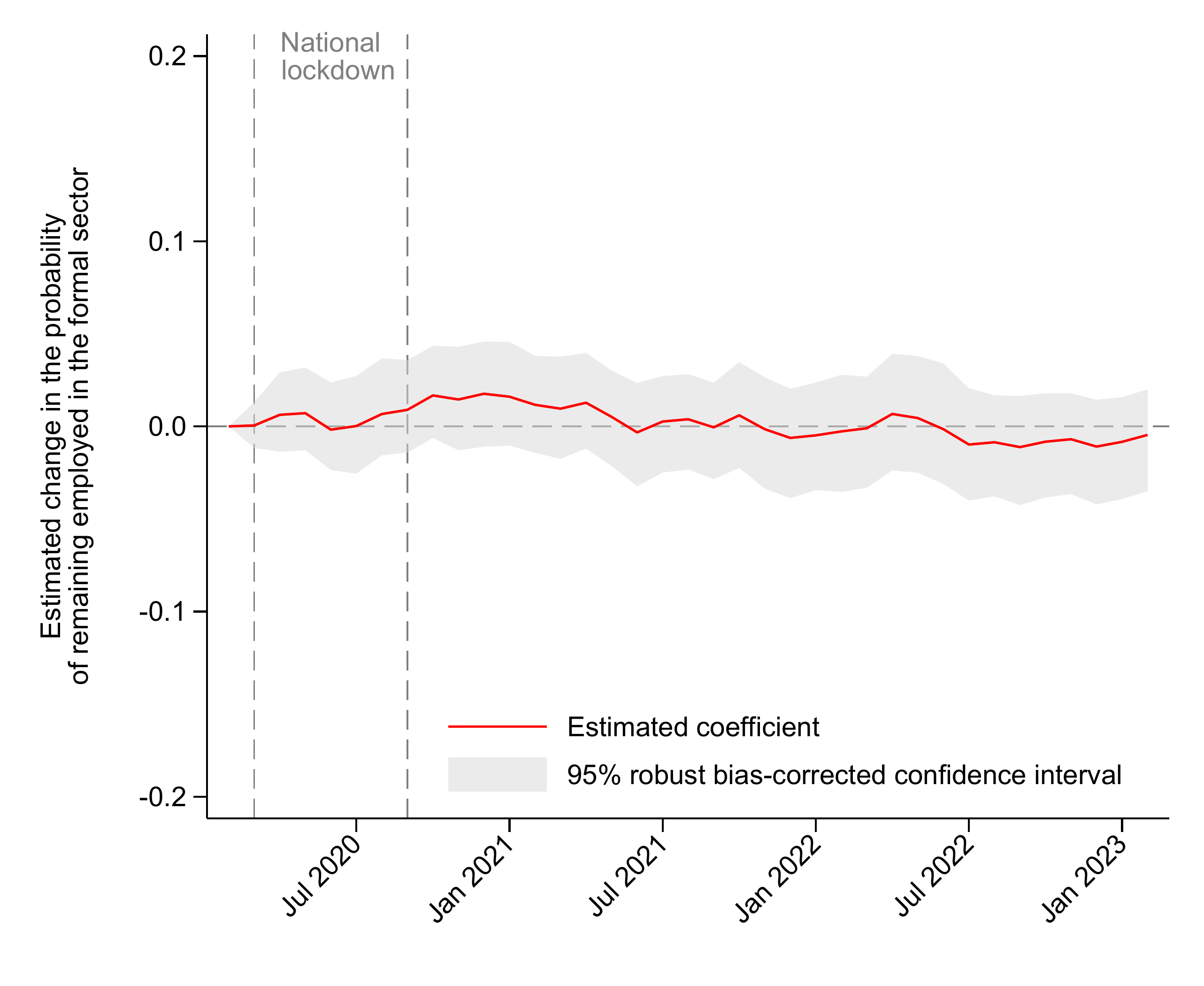}\\
	\noindent\makebox[\textwidth][c]{
		\noindent\begin{minipage}[c]{0.80\textwidth}	
			\justifying
			\noindent\textit{Notes}: The figure shows the estimated coefficients and the 95\% robust bias-corrected confidence intervals of the regression discontinuity exercises estimated separately for each month. Each analysis consists in local linear regressions (which allow for different slopes on either side of the cut-off) estimated using a triangular kernel and data-driven MSE-optimal bandwidths. All of them include the following covariates: child gender, child age, squared child age, child ethnicity, household head's educational attainment, household head's marital status, household size, area of residence (urban or rural) and canton fixed effects. Standard errors are clustered at the household level using robust inference.\\ 
			\noindent\textit{Source}: Authors' analysis from \textcite{socialregistry2025} and \textcite{socialsecurity2025}.
		\end{minipage}
	}
\end{figure}

To allow for better visualisation of the results, in Figure~\ref{Figure 7}, we show the estimated ITT effect month by month from March 2020 to February 2023 together with its 95\% confidence interval. Hereafter, we make use of this intuitive and compact format to summarise the impact of being eligible for the HDG in 2008/2009. Consistent with the previous results, we cannot reject that the estimated effect is not statistically different from zero at the 5\% level for any of the 36 months included in the analysis. 

Our results are also coherent with the virtual absence of overall long-term effects reported by \textcite{araujo2018} and \textcite{ponce2025}, who find no average impact on employment and labour market outcomes in the formal sector for the youngsters who received the HDG between 10--15 years earlier. Below, we discuss several potential explanations for our findings and their possible implications. 

Firstly, while the HDG seems to have a positive influence on the amount of education received \parencite{schady2008a,oosterbeek2008}, prior studies are generally more sceptical about its impact on educational quality. For instance, employing a research strategy similar to ours, \textcite{ponce2010} report no significant effect of either grant eligibility or actual receipt on maths and language test scores for children aged 5 to 16 years near the eligibility threshold during 2004–-2005. Using a randomised evaluation, \textcite{araujo2018} track children who were five or younger at the programme's start and assess their outcomes a decade later, finding no improvements in test scores due to early exposure to the HDG. This persistent lack of influence on educational quality may help to explain the null effect observed in later labour market outcomes for young adults.

Secondly, there are additional pathways through which the HDG could influence long-term labour market outcomes, although the evidence remains inconclusive. For example, studies on the programme's impact on cognitive development present mixed results \parencite{paxson2007,paxson2010,fernald2011,ponce2010}. Regarding health---another key aspect of human capital---the literature on the HDG is similarly cautious. Some research reports modest improvements in haemoglobin levels and access to deworming treatments and nutritional supplements for certain groups of children in rural areas \parencite{paxson2010,fernald2011}. \textcite{buser2017} observe that the loss of the HDG after seven years has a detrimental effect on children’s nutritional status, yet, somewhat unexpectedly, they find no clear benefits associated with gaining access to the grant. More recently, a non-experimental analysis using county-level panel data \parencite{moncayo2019} indicates that the HDG’s expansion significantly reduced under-five mortality---especially from poverty-related illnesses such as malnutrition, diarrhoeal diseases and lower respiratory tract infections.

The third explanation has to do with the nature and dynamics of the informal sector in Ecuador. If the dualistic perspective on informality holds, as recent surveys appear to suggest \parencite{laporta2014}, then increasing the workforce's educational attainment does not necessarily lead to a corresponding expansion of formal employment. According to this view, the formal and informal sectors operate in a largely disconnected manner, characterised by stark differences in productivity. Informality is seen as a consequence of the limited dynamism in the formal economy, which is regarded as the main driver of economic growth. The formal sector's limited capacity to absorb additional workers may stem from a range of structural factors, including patterns of international trade specialization \parencite{cimoli2006} and a shortage of highly educated entrepreneurs \parencite{laporta2014}.\footnote{The orthodox view of informality---where employment in this sector is seen as a voluntary choice by rational economic agents seeking to avoid the burdens and regulations of the formal economy---could also partly explain these null results. This perspective links informality to high intertemporal discount rates, which helps to explain why some workers choose to forgo social security coverage in favour of higher take-home pay \parencite{perry2007}. Indeed, high discount rates---believed to be transmitted within generations \parencite{knowles2023}---were assumed by the designers of CCTs to characterise their potential beneficiaries. However, the plausibility of this explanation is constrained by the clearly disadvantaged nature of Ecuador's informal sector.}

In a related argument, \textcite{pritchett2001} contends that increases in educational attainment do not necessarily translate into higher wages when the demand for educated labour remains stagnant. Such a situation may arise either because the skill intensity of the economy is not increasing, or because structural shifts---where more educated workers typically have an advantage due to their greater adaptability---are occurring too slowly. This line of reasoning can be extended to explain the limited availability of higher-quality jobs within the formal sector.

Finally, the lack of strict enforcement of conditionalities by the Ecuadorian authorities (they announced the requirements but never enforced them) may help to explain the programme's relatively modest impact on human capital accumulation, from school enrolment to health and other related domains. The line between unconditional and conditional transfers is blurry \parencite{baird2013}: social benefits can differ not only in their rules on the paper but also in the rules on the paper, the dissemination of information to the relevant population, the importance placed on of the value of schooling for children as communicated by the authorities, the monitoring of conditions and the enforcement of any penalties or sanctions. In this respect, it seems more accurate to characterise design options as falling somewhere on a continuum between pure unconditional cash transfers (UCTs) and heavy-handed CCTs. Previous research tends to support that conditionality exerts a causal effect on the short-term effectiveness of cash transfers \parencite{baird2013}. Therefore, it seems reasonable to think that long-term outcomes are also related to the enforcement of requirements.\footnote{According to the classification proposed by \textcite{baird2013}, the Ecuador's HDG ranks low in terms of conditionality compared to other CCT programmes in Latin America and the Caribbean, where previous research has sometimes identified a positive long-run impact. In particular, these authors distinguish seven types of programmes according to the criteria mentioned in the text: UCT programmes unrelated to children or education (0), UCT programmes targeted at children with an explicit aim of improving education (1), UCTs that are conducted within a rubric of education (2), CCTs with explicit conditions on paper or education encouragement, but not monitored or enforced (3), CCTs with explicit conditions, (imperfectly) monitored, with minimal enforcement (4), CCTs explicit conditions with monitoring and enforcement of enrolment condition (5) and CCTs with explicit conditions with monitoring and enforcement of attendance condition (6). They consider the Ecuadorian HDG as a CCT programme with explicit conditions on paper or education encouragement with no monitoring or enforcement (3). Within the region, the countries where previous studies have found more positive long-term outcomes tend to receive higher scores in terms of conditionality: the \textit{Programa de Asignaci{\'o}n Familiar II} in Honduras, 5; \textit{Red de Protecci{\'o}n Social} in Nicaragua, 6 and \textit{Progresa} in Mexico, 4. \textcite{baird2013} place Brazilian \textit{Bolsa Escola}, which seems to have a relevant positive effect in long-term labour market outcomes in the same category as the HDG and give its successor, \textit{Bolsa Familia}, a score of 4. \textit{Familias en acci{\'o}n} in Colombia, with a relatively low impact in the long run, receives a score of 6.} Although we do not have any specific experimental or quasi-experimental evidence for the Ecuadorian case, the positive impact on educational enrolment reported in a previous study appears to be relevant only for those households who believe that the government is effectively monitoring the fulfilment of the conditions \parencite{schady2008a}.

Regarding the external validity of our findings, we should bear in mind that we estimate only the local ITT effects of the HDG. In other words, our results only apply to the neighbourhood of the specific cut-off point in 2008/2009. It is possible that we could observe better results for households with a lower socio-economic status. In fact, previous work has revealed that children living in the country's poorest households are those who experience the largest short-run impacts (in terms of school enrolment and health) of the HDG \parencite{paxson2010,oosterbeek2008,schady2008a}.\footnote{As we are only able to leverage information corresponding to a poverty index between 30.5 and 42.5, we cannot extrapolate ITT effects \enquote{away from the cut-off point}.}

\FloatBarrier
\subsection{Heterogeneity in the effects of the programme}\label{Subsection 4.2}

\begin{landscape}
	\begin{figure}[p]
		\scriptsize
		\captionsetup{width=1\linewidth}
		\caption{ITT effect heterogeneity}
		\label{Figure 8}
		\centering 
		\includegraphics[trim={0 0.5cm 0.5cm 0}, width=1\linewidth]{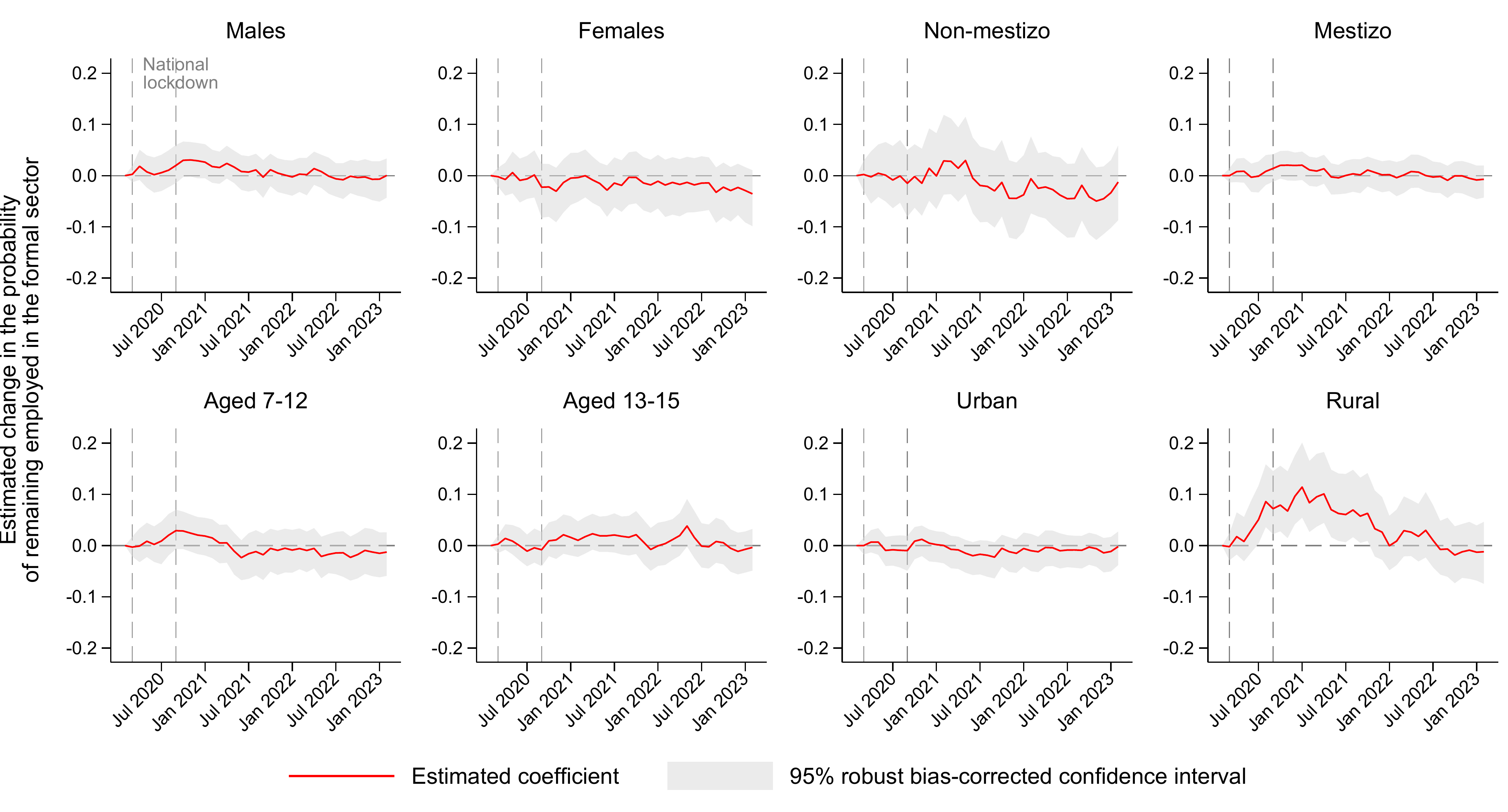}\\
		\noindent\makebox[\textwidth][c]{
			\noindent\begin{minipage}[c]{1\linewidth}	
				\justifying
				\noindent\textit{Notes}: The figures show the estimated coefficients and the 95\% robust bias-corrected confidence intervals of regression discontinuity exercises estimated separately for each month. Each analysis consists in local linear regressions (which allow for different slopes on either side of the cut-off) estimated using a triangular kernel and data-driven MSE-optimal bandwidths. All of them include the following covariates: child gender, child age, squared child age, child ethnicity, household head's educational attainment, household head's marital status, household size and area of residence (urban or rural)---excluding the covariate associated to each group appropriately---and canton fixed effects. Standard errors are clustered at the household level using robust inference.\\ 
				\noindent\textit{Source}: Authors' analysis from \textcite{socialregistry2025} and \textcite{socialsecurity2025}.
			\end{minipage}
		}
	\end{figure}
\end{landscape}

Since the literature shows that CCTs---including the HDG---have different effects across population groups, this subsection explores whether the impact of the grant on resilience to the COVID-19 crisis differed by sex, ethnicity, age at the time of potential exposure to the transfer and area of residence in 2008/2009 (Figure~\ref{Figure 8}). We find no impact on males, females, mestizos, non-mestizos, children aged 7--12 years, children aged 13--15 years or persons living in urban areas. Nevertheless, we clearly observe a significant impact on young people living in rural areas at the time of eligibility for the grant. The estimated effects are substantial: in particular, the share of people employed in the formal sector was approximately 5--10 percentage points higher among those who were eligible for the grant in 2008/2009 from the final months of the national lockdown until the end of 2021.   

Yet, this finding is not at odds with previous literature on the HDG. In fact, several works identify some positive short-run effects of receiving the grant on child development exclusively in rural areas (and not observed in urban ones). For instance, \textcite{paxson2010} find a significant impact of the HDG on both cognitive and behavioural (specifically, long-term memory) and physical measures (particularly, higher haemoglobin levels), as well as a higher likelihood of receiving deworming treatments. Similarly, \textcite{fernald2011} report a positive effect on language skills and a larger probability of getting vitamin A or iron supplementation. Finally, the work of \textcite{schady2008b} suggests that, after the intervention, eligible rural households had significantly higher food shares than non-eligible ones. 

Interestingly, the pattern of resilience observed among eligible individuals living in rural areas at the time of eligibility does not obey to differences in the sector of activity in February 2020 (which would be an outcome of the HDG). Figure~\ref{Figure A3} shows the ITT effect conditional on pre-pandemic industry for this group, which is virtually the same as that presented in Figure~\ref{Figure 8}.

It is informative to calculate the probability of having remained employed for a representative individual in a rural area in 2008/2009 (at the cut-off), being eligible and ineligible for the HDG in childhood (instead of computing just the difference between an eligible and ineligible person). This approach provides details on how exactly the HDG protected its recipients during the pandemic (e.g., whether it raised participation on formal employment or attenuated its decline). 

\begin{figure}[p]
	\footnotesize
	\centering
	\captionsetup{width=0.80\textwidth}
	\caption{Probability of remaining employed by eligibility status for individuals living in rural areas in 2008/2009}
	\label{Figure 9}
	\centering 
	\includegraphics[width=0.80\textwidth]{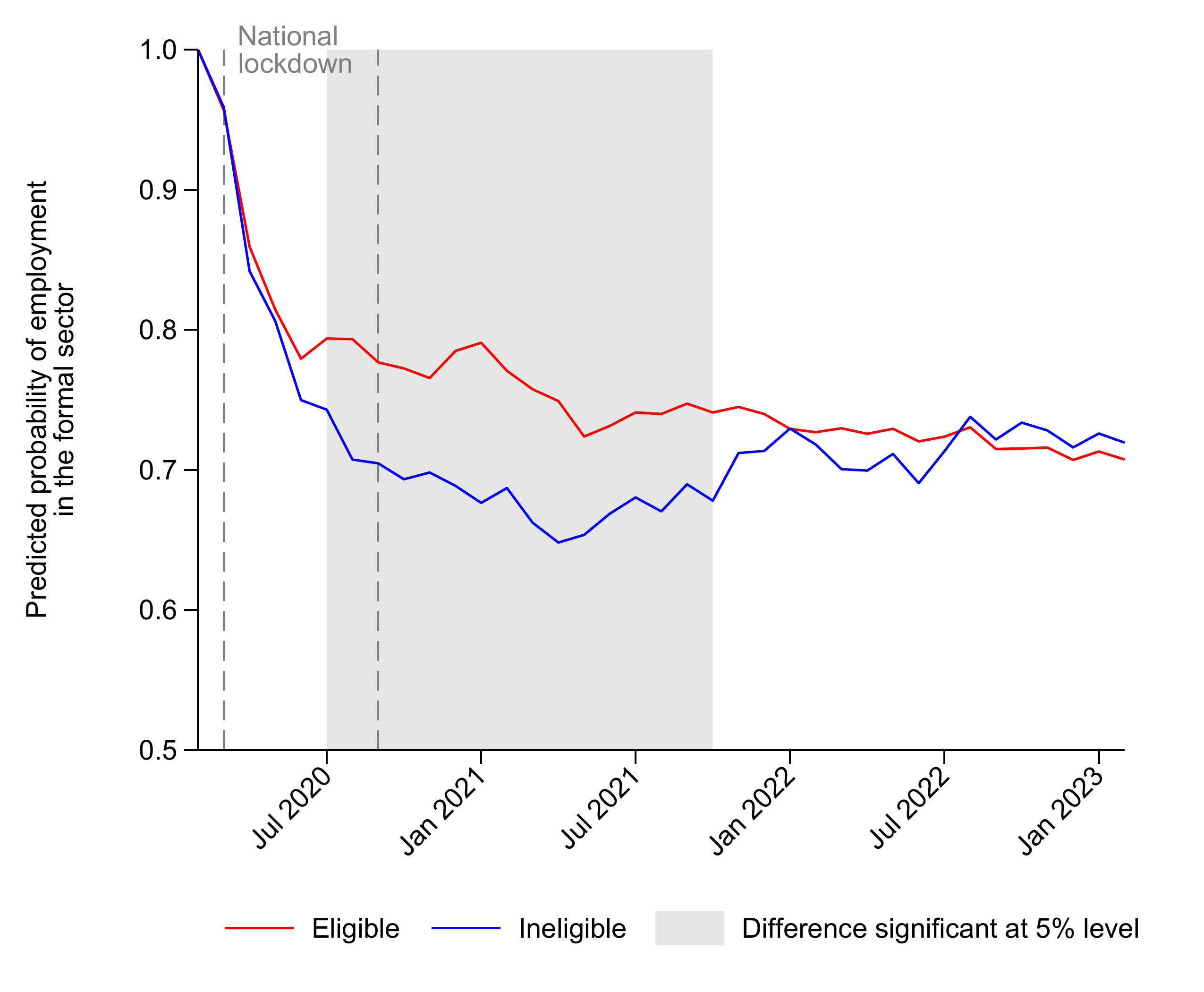}\\
	\noindent\makebox[\textwidth][c]{
		\noindent\begin{minipage}[c]{0.80\textwidth}	
			\justifying
			\noindent\textit{Notes:} The figure shows the probability of remaining employed by month using the results of the RDD presented in Figure~\ref{Figure 8} evaluated for an individual in a rural area in 2007/2009 at the cut-off with all control covariates set to their median values. \\
			\noindent\textit{Source}: Authors' analysis from \textcite{socialregistry2025} and \textcite{socialsecurity2025}.
		\end{minipage}
	}
\end{figure}

In this respect, Figure~\ref{Figure 9} shows the probability of having remained employed in the formal economy for an individual in a rural area in 2008/2009, with all control covariates set to their median values. This likelihood was retrieved from estimating Equation~\ref{Equation 1} separately for eligible and ineligible persons living in a rural environment during their childhood. In particular, the red line in the graph depicts the probability of remaining employed for an eligible person in a rural area in 2008/2009 by month. Such a likelihood is given by the expression $\hat{\alpha} + \hat{\tau} + \bar{Z}_i' \hat{\Theta}$, where $\bar{Z}_i$ consists in a vector of covariates with all control variables set to their median values. The blue line shows the same probability for an ineligible individual, derived from $\hat{\alpha} + \bar{Z}_i' \hat{\Theta}$. 

We observe a marked decrease in participation in formal employment for both groups during the national lockdown. For eligible individuals, this decline slowed significantly from July 2020 onwards, while it continued until May 2021 for the ineligible population. The latter group then recovered, catching up with the former by November 2021. No statistically significant differences was observed between the two groups thereafter.

The results observed for youngsters who lived in rural areas in 2008/2009 make it interesting to assess how the patterns of participation in the formal sector vary across different socio-economic characteristics. In Figure~\ref{Figure A4}, we show that such an outcome was driven entirely by males. Differences by age were mostly negligible and only mestizos exhibited this resilience pattern, partly due to the reduced non-mestizo sample which leads to quite imprecise estimates.  

These differences by gender may be due to two factors. Firstly, females made up the majority of the workforce in more contact-intensive services particularly hit by the pandemic crisis, such as wholesale and retail trade, accommodation and food services, education and other personal service activities \parencite{armijos2023}. In these sectors, the possibility of working from home or making other adaptations to the pandemic situation was less feasible. 

Secondly, school closures (prolonged until May 2022) had a much more negative impact on the women's labour market outcomes than on males'. As in other countries \parencite{russell2020,zamarro2021}, mandated school closures disproportionately affected unskilled females with children \parencite{imf2021}. These women had to assume more care responsibilities, which limited their ability to adapt to the new labour market trends. Overall, these results align with the literature highlighting the less effective role of human capital in protecting women from economic downturns \parencite{cappelli2021}.

\subsection{Robustness checks}\label{Subsection 4.3}

In this subsection, we present the results of several robustness checks performed to assess the stability of our main findings. First, we employ multivariate imputation techniques to address the issue of missing labour market outcomes (due to the missing identity card numbers) in accordance to the procedure outlined in Subsection~\ref{Subsection 3.2}. Second, we restrict our analysis to the complete case sample. Third, we assess the impact of selecting a CER-optimal bandwidth. Fourth, we employ local quadratic regressions. Fifth, we repeat the analysis excluding those observations with a poverty index within 0.1 points of the cut-off (a \enquote{donut hole} RDD). Sixth, as mentioned above, we estimate a fuzzy RDD instrumenting HDG receipt in 2010 with the discontinuity around the threshold. Finally, we perform two falsification or placebo tests. Following the suggestion of \textcite{imbens2008}, we test for jumps at the median of the two subsamples on either side of the cut-off value. Since these are non-discontinuity points, we should not reject the null hypothesis that the effect is absent in those instances.

\begin{landscape}
	\begin{figure}[p]
		\scriptsize
		\captionsetup{width=1\linewidth}
		\caption{Robustness checks (ITT and LATE)}
		\label{Figure 10}
		\centering 
		\includegraphics[trim={0 0.5cm 0.5cm 0}, width=1\linewidth]{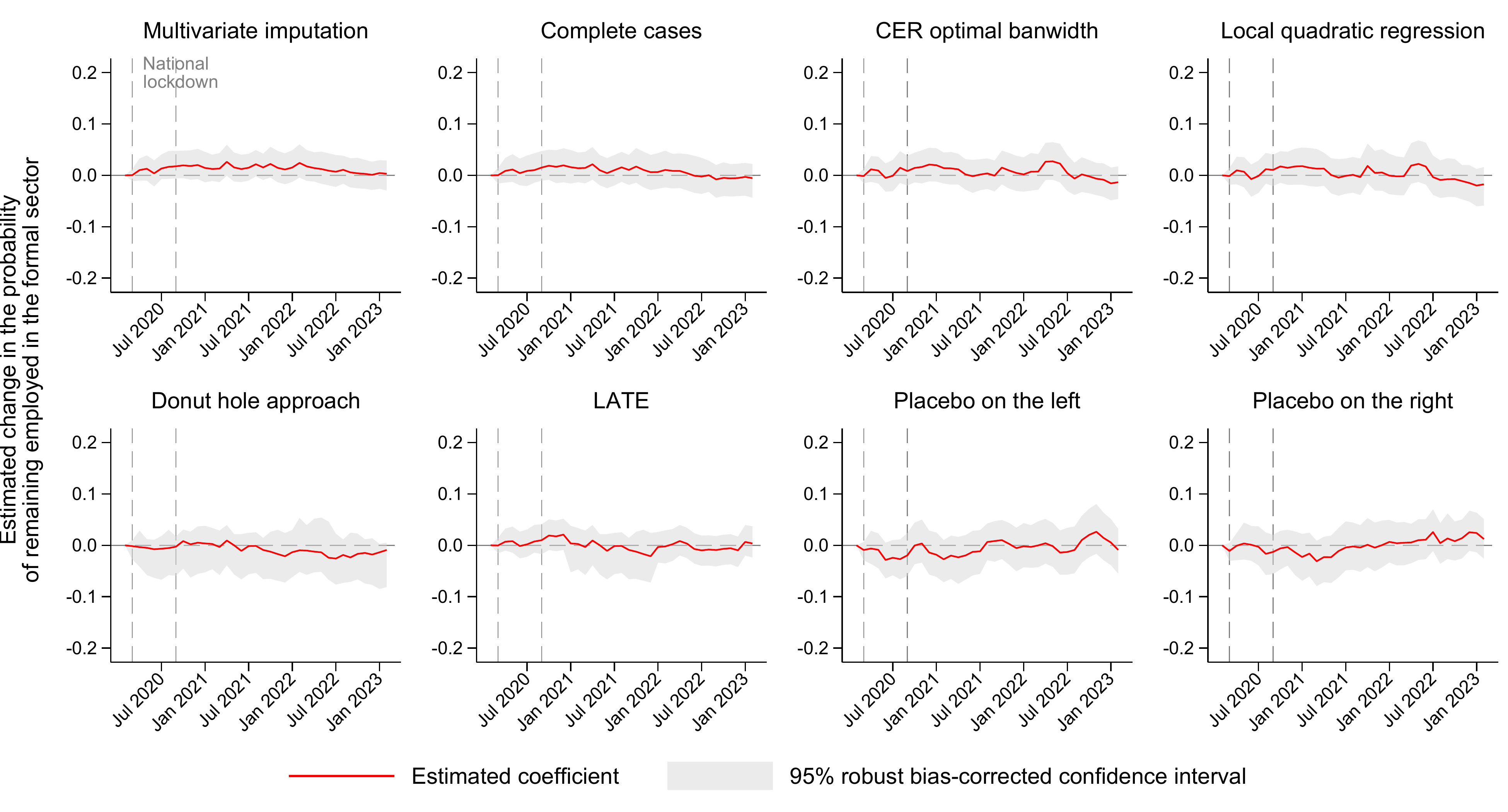}\\
		\noindent\makebox[\textwidth][c]{
			\noindent\begin{minipage}[c]{1\linewidth}	
				\justifying
				\noindent\textit{Notes}: The figures show the estimated coefficients and the 95\% robust bias-corrected confidence intervals of regression discontinuity exercises estimated separately for each month (whose details are provided in the text). All of them include the following covariates: child gender, child age, squared child age, child ethnicity, household head's educational attainment, household head's marital status, household size and canton fixed effects. Standard errors are clustered at the household level using robust inference.\\ 
				\noindent\textit{Source}: Authors' analysis from \textcite{socialregistry2025} and \textcite{socialsecurity2025}.
			\end{minipage}
		}
	\end{figure}
\end{landscape}

The results of these sensitivity analyses, presented in Figure~\ref{Figure 10}, are encouraging. They indicate that the six alternative estimation strategies outlined above do not affect our conclusions. The \textit{F}-statistic of the first stage in the fuzzy design is above \num{10000} in all cases. The upper panel of Figure~\ref{Figure A5} shows the plot of the first stage for March 2020 by way of example. This almost perfect compliance applies to all the months considered in the analysis. Furthermore, the two falsification tests suggest the absence of relevant discontinuities at the two points proposed above, reaffirming our confidence in our identification strategy.

\begin{landscape}
	\begin{figure}[p]
		\scriptsize
		\captionsetup{width=1\linewidth}
		\caption{Robustness checks for individuals in rural areas in 2008/2009 (ITT and LATE)}
		\label{Figure 11}
		\centering 
		\includegraphics[trim={0 0.5cm 0.5cm 0}, width=1\linewidth]{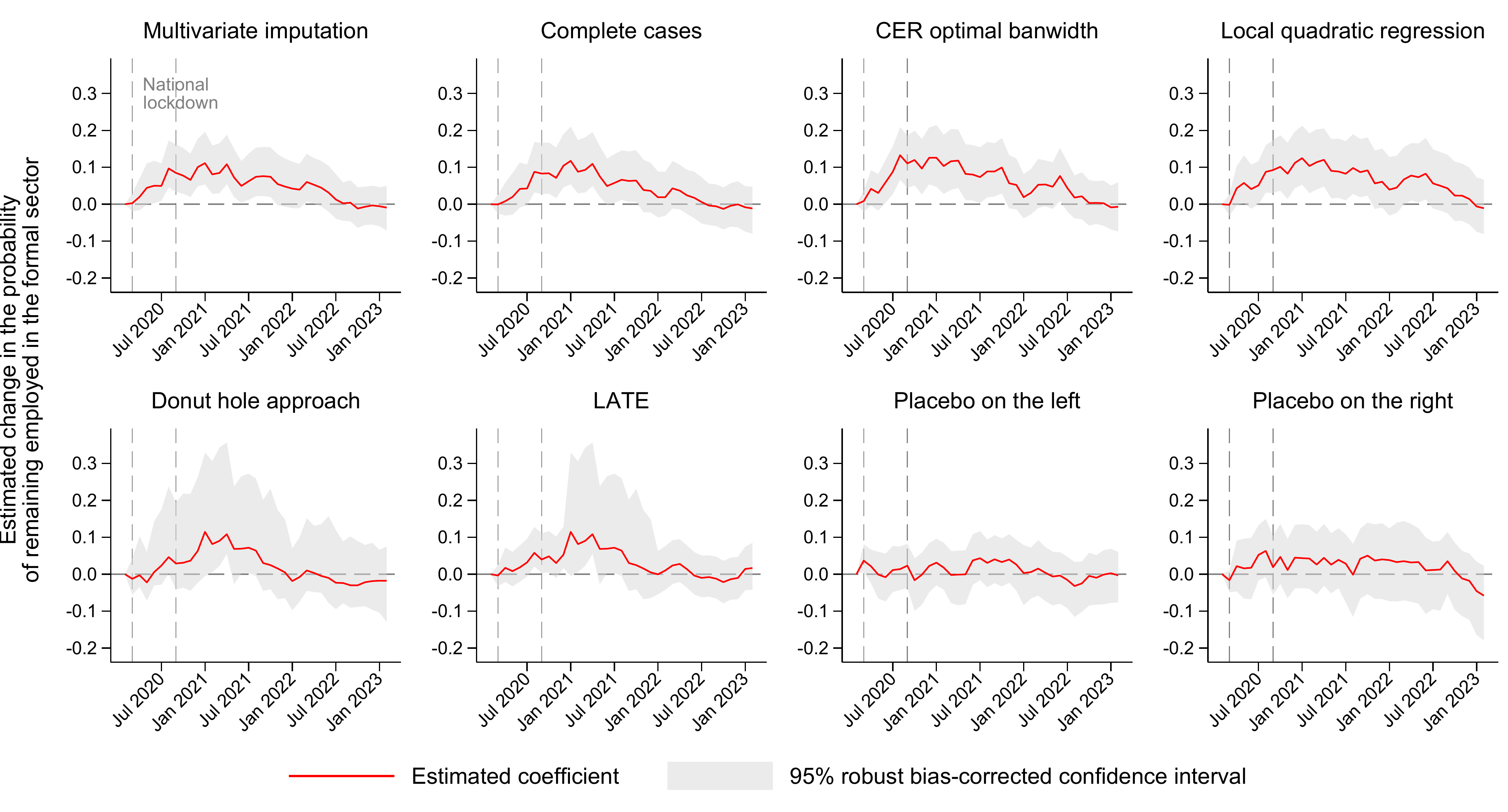}\\
		\noindent\makebox[\textwidth][c]{
			\noindent\begin{minipage}[c]{1\linewidth}	
				\justifying
				\noindent\textit{Notes}: The figures show the estimated coefficients and the 95\% robust bias-corrected confidence intervals of regression discontinuity exercises estimated separately for each month (whose details are provided in the text). All of them include the following covariates: child gender, child age, squared child age, child ethnicity, household head's educational attainment, household head's marital status, household size and canton fixed effects. Standard errors are clustered at the household level using robust inference.\\ 
				\noindent\textit{Source}: Authors' analysis from \textcite{socialregistry2025} and \textcite{socialsecurity2025}.
			\end{minipage}
		}
	\end{figure}
\end{landscape}

We repeat of all the battery of stability tests on the population living in rural areas when tehy were potentially eligible for the HDG (Figure~\ref{Figure 11}). The six distinct versions of the RDD convey the same message as our main results: these workers were more likely to remain employed during the hardest times of the pandemic. The \textit{F}-statistic of the first stage of the fuzzy discontinuity design exceeds \num{10000} in all cases. The graphical illustration of the first stage is shown in the lower panel of Figure~\ref{Figure A5}. Similarly, the two placebo tests are reassuring, since they indicate that there is no statistically significant discontinuity at the two points considered in these falsification exercises.

\FloatBarrier
\section{Conclusion}\label{Section 5}

Despite having become flagship social programmes in several countries, empirical evidence on the long-term impacts of CCTs remains scarce, and our understanding of their potential to help individuals cope with shocks later in life is still limited. This paper aims to fill this gap in the literature and shed light on this issue by leveraging data from the HDG in Ecuador---one of the pioneering programmes of its kind---during the COVID-19 pandemic.

Although previous research indicates that the HDG has a positive effect on school enrolment, our findings suggest that eligibility for this grant has no overall impact on resilience---which we understand as the ability to have remained employed in the formal sector during the pandemic---is null. We attribute these results primarily to the programme' weak conditionality, particularly compared to other CCTs in the region, to the limited capacity of the formal economy to absorb even a more educated labour force and relatively meagre short-term results of the grant beyond children's participation in schooling in comparison with similar schemes in Latin America and the Caribbean. Additionally, as we only provide estimates of the local ITT effects, we cannot rule out the possibility that this CCT has an impact away from the discontinuity we have exploited in our analysis.

Nevertheless, it is worth noting that we have found that the HDG increased pandemic resilience among individuals living in rural areas when they became eligible for the transfer, roughly 12 years prior to the outbreak. We have interpreted this finding as being due to the significantly larger short-term impacts on child development reported by previous studies.

In light of existing evidence on the long-run impacts of CCTs and the role of conditionality in enhancing their effectiveness, we believe that strengthening the enforcement of conditions---so that the HDG can generate a greater short-term impact---may help to improve the programme's long-term outcomes, including its capacity to enhance recipients' resilience to shocks later in life.

\clearpage
\singlespacing
\printbibliography
\clearpage

% Figure 1. Excess mortality from the COVID-19 in Ecuador (%, 1st January 2020-30th December 2023)
\clearpage

% Figure 2. Ecuador government response to COVID-19
\clearpage

% Figure 3. Evolution of employment in Ecuador (December 2019-December 2024, December 2019 = 100)
\clearpage

% Figure 4. Proportion of individuals employed in the formal sector (February 2020 to February 2022)
\clearpage

% Figure 5. Test for manipulation of the assignment variable based on density discontinuity
\clearpage

% Figure 6. Graphical analysis
\clearpage

% Figure 7. Graphical analysis
\clearpage

% Figure 8. Effect heterogeneity
\clearpage

% Figure 9. Probability of remaining employed by eligibility status
\clearpage

% Figure 10. Robustness checks
\clearpage

% Figure 11. Robustness checks
\clearpage

% Table 1. Descriptive statistics
\begin{center}
	\begin{singlespace}
		\begin{table}[p]
			\centering
			\begin{threeparttable}
				\centering
				\def\sym#1{\ifmmode^{#1}\else\(^{#1}\)\fi}
				\footnotesize
				\setlength\tabcolsep{3pt}
				\sisetup{
					table-number-alignment = center,
				}
				\begin{TableNotes}[flushleft]\setlength\labelsep{0pt}\footnotesize\justifying
					\item\textit{Source}: Authors' analysis from \textcite{socialregistry2025} and \textcite{socialsecurity2025}.
				\end{TableNotes}
				\begin{tabularx}{0.8\textwidth}{X *{2}{S[table-column-width=1.7cm]}}
					\caption{Summary statistics} \label{Table 1} \\		
					\toprule
					&\multicolumn{1}{c}{Mean}&\multicolumn{1}{c}{\makecell{Standard\\deviation}}\\
					\midrule
					Employed in February 2020&1.000&0.000\\
Employed in June 2020&0.797&0.402\\
Employed in December 2020&0.744&0.436\\
Employed in June 2021&0.715&0.451\\
Employed in December 2021&0.708&0.455\\
Employed in June 2022&0.692&0.462\\
Employed in December 2022&0.682&0.466\\
Poverty index in 2008/2009&36.345&3.421\\
HDG recipiency in 2010&0.559&0.497\\
Female in 2008/2009&0.353&0.478\\
Mestizo in 2008/2009&0.818&0.386\\
Age in 2008/2009&12.325&2.030\\
Household head's years of education in 2008/2009&7.165&3.369\\
Household head married or in union in 2008/2009&0.642&0.479\\
Household size in 2008/2009&4.775&1.572\\
Rural area in 2008/2009&0.255&0.436\\ [1ex]
No. of observations&\multicolumn{1}{c}{\hspace{1.5mm}\num{80152}}&
\\
					\bottomrule
					\insertTableNotes
				\end{tabularx}
			\end{threeparttable}
		\end{table}
	\end{singlespace}
\end{center}
\clearpage

% Table 2. Covariate balance: evaluation of the discontinuity in the covariates at the cut-off
\begin{landscape}
	\begin{singlespace}
		\begin{table}[p]
			\begin{ThreePartTable}
				\centering
				\def\sym#1{\ifmmode^{#1}\else\(^{#1}\)\fi}
				\footnotesize
				\setlength\tabcolsep{4pt}
				\sisetup{
					table-number-alignment = center,
				}
				\begin{TableNotes}[flushleft]\setlength\labelsep{0pt}\footnotesize\justifying
					\item\textit{Notes}: \sym{***} significant at the 1\% level; \sym{**} significant at the 5\% level; \sym{*} significant at the 10\% level. The table shows the estimates from local linear polynomial regressions (which allow for different slopes on either side of the cutoff) estimated using a triangular kernel and data-driven MSE-optimal bandwidths. Statistical significance is based on robust-bias corrected inference. All specifications include canton fixed effects. Standard errors clustered at the household level are in parentheses.
					\item\textit{Source}: Authors' analysis from \textcite{socialregistry2025} and \textcite{socialsecurity2025}.
				\end{TableNotes}
				\begin{tabularx}{\linewidth}{X *{7}{S[table-column-width=2cm]}}
					\caption{Covariate balance: evaluation of the discontinuity in the covariates at the cut-off} \label{Table 2} \\		
					\toprule
					&\multicolumn{1}{c}{(I)}&\multicolumn{1}{c}{(II)}&\multicolumn{1}{c}{(III)}&\multicolumn{1}{c}{(IV)}&\multicolumn{1}{c}{(V)}&\multicolumn{1}{c}{(VI)}&\multicolumn{1}{c}{(VII)}\\ [6pt]					
					&\multicolumn{1}{c}{\makecell{Female}}&
					\multicolumn{1}{c}{\makecell{Mestizo}}&
					\multicolumn{1}{c}{\makecell{Age}}&
					\multicolumn{1}{c}{\makecell{Household\\head's years\\of education}}&\multicolumn{1}{c}{\makecell{Household\\head's married\\or in union}}&
					\multicolumn{1}{c}{\makecell{Household\\size}}&
					\multicolumn{1}{c}{\makecell{Rural\\area}}\\
					\midrule
					ITT effect&-0.008&-0.007&0.066&-0.054&0.005&-0.003&0.014\\
&(0.014)&(0.010)&(0.059)&(0.093)&(0.016)&(0.052)&(0.011)\\ [1ex]
No. of observations &\multicolumn{1}{c}{\hspace{1.5mm}\num{80152}}&\multicolumn{1}{c}{\hspace{1.5mm}\num{80152}}&\multicolumn{1}{c}{\hspace{1.5mm}\num{80152}}&\multicolumn{1}{c}{\hspace{1.5mm}\num{80152}}&\multicolumn{1}{c}{\hspace{1.5mm}\num{80152}}&\multicolumn{1}{c}{\hspace{1.5mm}\num{80152}}&\multicolumn{1}{c}{\hspace{1.5mm}\num{80152}}\\
No. of observations effectively used&\multicolumn{1}{c}{\hspace{1.5mm}\num{21348}}&\multicolumn{1}{c}{\hspace{1.5mm}\num{30582}}&\multicolumn{1}{c}{\hspace{1.5mm}\num{21855}}&\multicolumn{1}{c}{\hspace{1.5mm}\num{27214}}&\multicolumn{1}{c}{\hspace{1.5mm}\num{18556}}&\multicolumn{1}{c}{\hspace{1.5mm}\num{21899}}&\multicolumn{1}{c}{\hspace{1.5mm}\num{25765}}\\
Mean of dependent variable&0.353&0.818&12.325&7.165&0.642&4.775&0.255\\
Standard deviation of dependent variable&0.478&0.386&2.030&3.369&0.479&1.572&0.436
\\
					\bottomrule
					\insertTableNotes
				\end{tabularx}
			\end{ThreePartTable}
		\end{table}
	\end{singlespace}
\end{landscape}
\clearpage

% Table 3. Main results
\begin{landscape}
	\begin{singlespace}
		\begin{table}[p]
			\begin{ThreePartTable}
				\def\sym#1{\ifmmode^{#1}\else\(^{#1}\)\fi}
				\footnotesize
				\setlength\tabcolsep{1pt}
				\sisetup{
					table-number-alignment = center,
				}
				\begin{TableNotes}[flushleft]\setlength\labelsep{0pt}\footnotesize\justifying
					\item\textit{Notes}: \sym{***} significant at 1\% level; \sym{**} significant at 5\% level; \sym{*} significant at 10\% level. The table shows the estimates from local linear regressions (which allow for different slopes on either side of the cutoff) estimated using a triangular kernel and data-driven MSE-optimal bandwidths. The control variables included in the second column are child gender, child age, squared child age, child ethnicity, household head's educational attainment, household head's marital status, household size and area of residence (urban or rural). Standard errors clustered at the household level are in parentheses.
					\item\textit{Source}: Authors' analysis from \textcite{socialregistry2025} and \textcite{socialsecurity2025}.				
				\end{TableNotes}
				\begin{tabularx}{\linewidth}{X *{6}{S[table-column-width=2.75cm]}}
					\caption{ITT effects on the probability of remaining employed in the formal sector in selected months} \label{Table 3}\\
					\toprule
					&\multicolumn{1}{c}{(I)}&\multicolumn{1}{c}{(II)}&\multicolumn{1}{c}{(III)}&\multicolumn{1}{c}{(V)}&\multicolumn{1}{c}{(V)}&\multicolumn{1}{c}{(VI)}\\
					&\multicolumn{1}{c}{Jun 2020}&\multicolumn{1}{c}{Dec 2020}&\multicolumn{1}{c}{Jun 2021}	&\multicolumn{1}{c}{Dec 2021}&\multicolumn{1}{c}{Jun 2022}&\multicolumn{1}{c}{Dec 2022}\\
					\midrule
					ITT effect&-0.002&0.018&-0.003&-0.006&-0.002&-0.011\\
&(0.010)&(0.012)&(0.012)&(0.013)&(0.014)&(0.012)\\ [1ex]
No. of observations&\multicolumn{1}{c}{\hspace{1.5mm}\num{80152}}&\multicolumn{1}{c}{\hspace{1.5mm}\num{80152}}&\multicolumn{1}{c}{\hspace{1.5mm}\num{80152}}&\multicolumn{1}{c}{\hspace{1.5mm}\num{80152}}&\multicolumn{1}{c}{\hspace{1.5mm}\num{80152}}&\multicolumn{1}{c}{\hspace{1.5mm}\num{80152}}\\
No. of observations effectively used&\multicolumn{1}{c}{\hspace{1.5mm}\num{29000}}&\multicolumn{1}{c}{\hspace{1.5mm}\num{24712}}&\multicolumn{1}{c}{\hspace{1.5mm}\num{27115}}&\multicolumn{1}{c}{\hspace{1.5mm}\num{24751}}&\multicolumn{1}{c}{\hspace{1.5mm}\num{20277}}&\multicolumn{1}{c}{\hspace{1.5mm}\num{27501}}\\ [1ex]
Mean of dependent variable&0.797&0.744&0.715&0.708&0.692&0.682\\
Standard deviation of dependent variable&0.402&0.436&0.451&0.455&0.462&0.466\\[1ex]
Control variables&\multicolumn{1}{c}{\checkmark}&\multicolumn{1}{c}{\checkmark}&\multicolumn{1}{c}{\checkmark}&\multicolumn{1}{c}{\checkmark}&\multicolumn{1}{c}{\checkmark}&\multicolumn{1}{c}{\checkmark}\\
Canton fixed effects&\multicolumn{1}{c}{\checkmark}&\multicolumn{1}{c}{\checkmark}&\multicolumn{1}{c}{\checkmark}&\multicolumn{1}{c}{\checkmark}&\multicolumn{1}{c}{\checkmark}&\multicolumn{1}{c}{\checkmark}
\\
					\bottomrule	
					\insertTableNotes
				\end{tabularx}
			\end{ThreePartTable}
		\end{table}
	\end{singlespace}
\end{landscape}
\clearpage

\setcounter{table}{0}
\setcounter{figure}{0}
\setcounter{page}{1}
\renewcommand\thetable{A\arabic{table}}
\renewcommand\thefigure{A\arabic{figure}}
\renewcommand\thepage{A\arabic{page}}

\appendix
\section*{Appendix}

\vspace*{\fill}
% Figure A1. Incidence of the COVID-19 pandemic in Ecuador (January 2020-February 2025)
\begin{figure}[ht]
	\footnotesize
	\captionsetup{width=0.65\textwidth}
	\caption{Incidence of the COVID-19 pandemic in Ecuador}
	\centering 
	\medskip
	\begin{subfigure}[p]{1\textwidth}
		\centering
		\captionsetup{width=0.65\textwidth}
		\caption{Weekly COVID-19 cases}
		\includegraphics[width=0.60\textwidth]{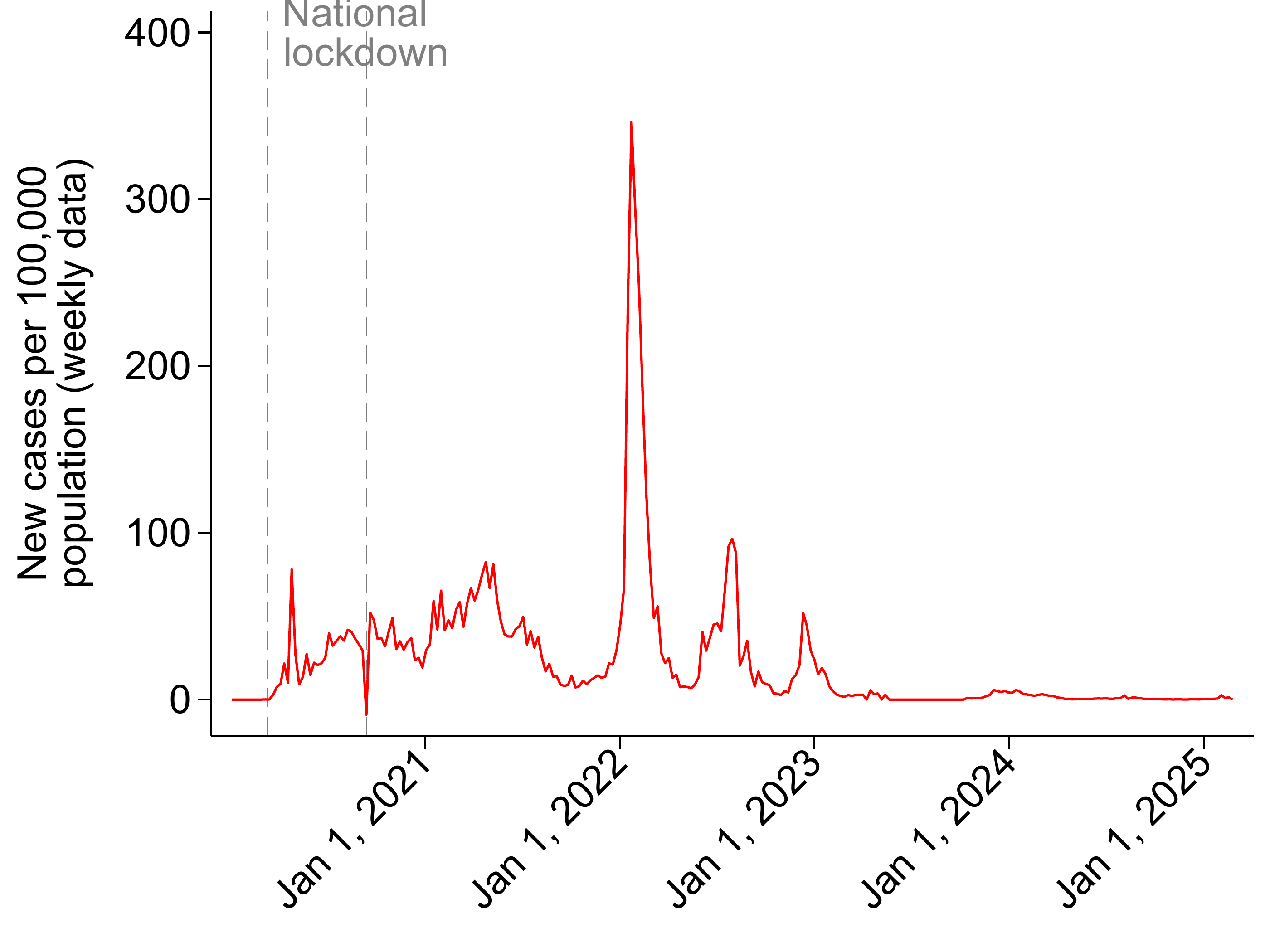} \\
	\end{subfigure}
	\begin{subfigure}[t]{1\textwidth}
		\centering
		\captionsetup{width=0.65\textwidth}
		\caption{Weekly COVID-19 deaths}
		\includegraphics[width=0.60\textwidth]{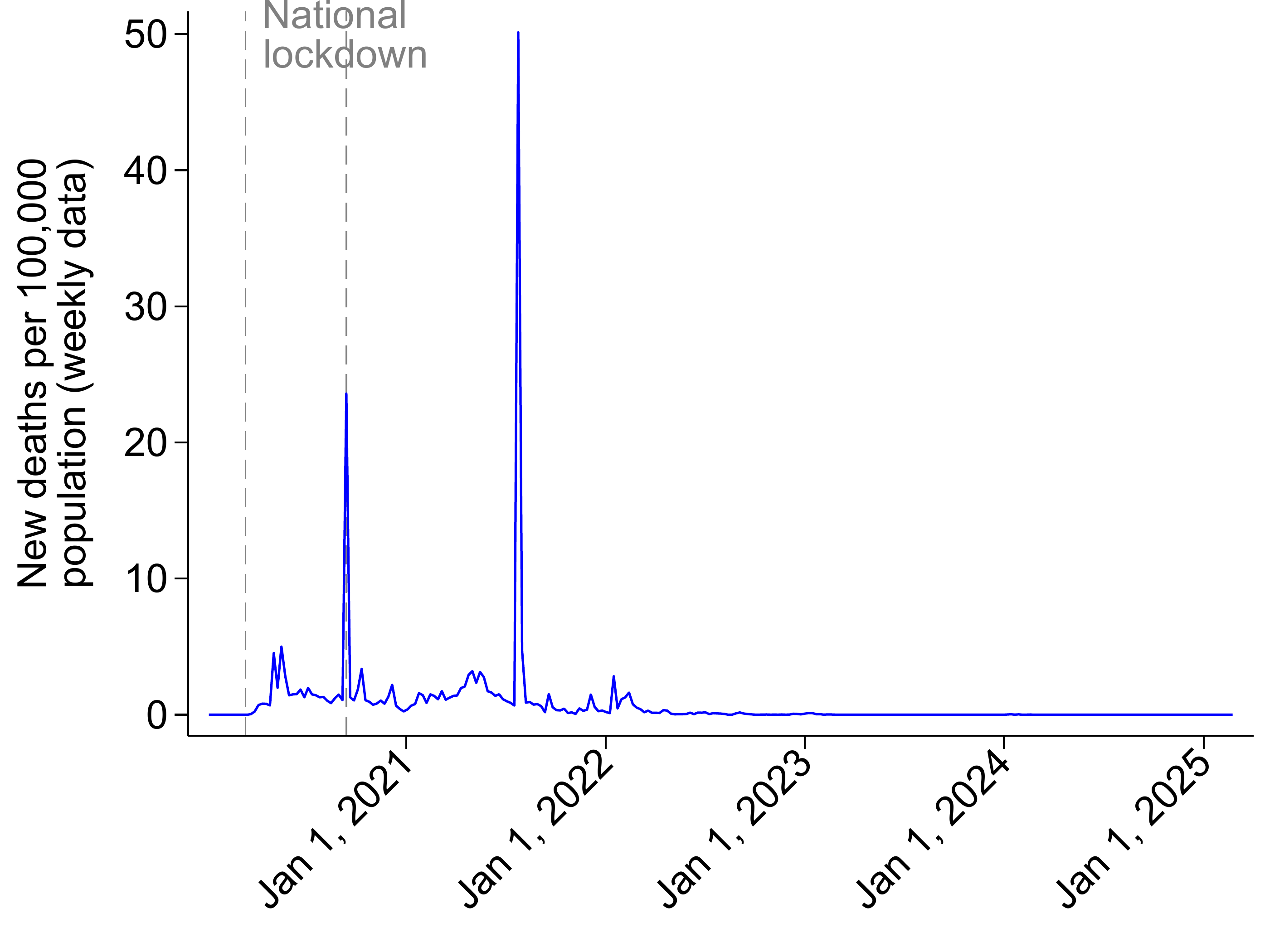} \\
	\end{subfigure}
	\noindent\makebox[\textwidth][c]{
		\noindent\begin{minipage}[c]{0.65\textwidth}	
			\justifying
			\noindent\textit{Note}: The two largest peaks in the death data correspond to the adjustments of 2020 and 2021 data performed by the government.\\ 
			\noindent\textit{Source}: Authors' analysis from \textcite{who2025}.
		\end{minipage}
	}
	\label{Figure A1}
\end{figure}
\vspace*{\fill}	

% Figure A2. Effects of the discontinuity on the proportion of individuals with missing identity card number and test for density continuity in the assignment variable including only individuals with identity card number
\begin{figure}[p]
	\scriptsize
	\captionsetup{width=0.8\textwidth}
	\caption{Effects of the discontinuity on the proportion of individuals with missing identity card number and test for density continuity in the assignment variable including only individuals with identity card number}
	\label{Figure A2}
	\centering 
	\medskip
	\begin{subfigure}[t]{1\textwidth}
		\centering
		\captionsetup{width=0.8\textwidth}
		\caption{Effects of the discontinuity on the proportion of individuals with missing identity card number}
		\includegraphics[width=0.7\textwidth, trim = 0cm 0cm 0cm 0cm, clip]{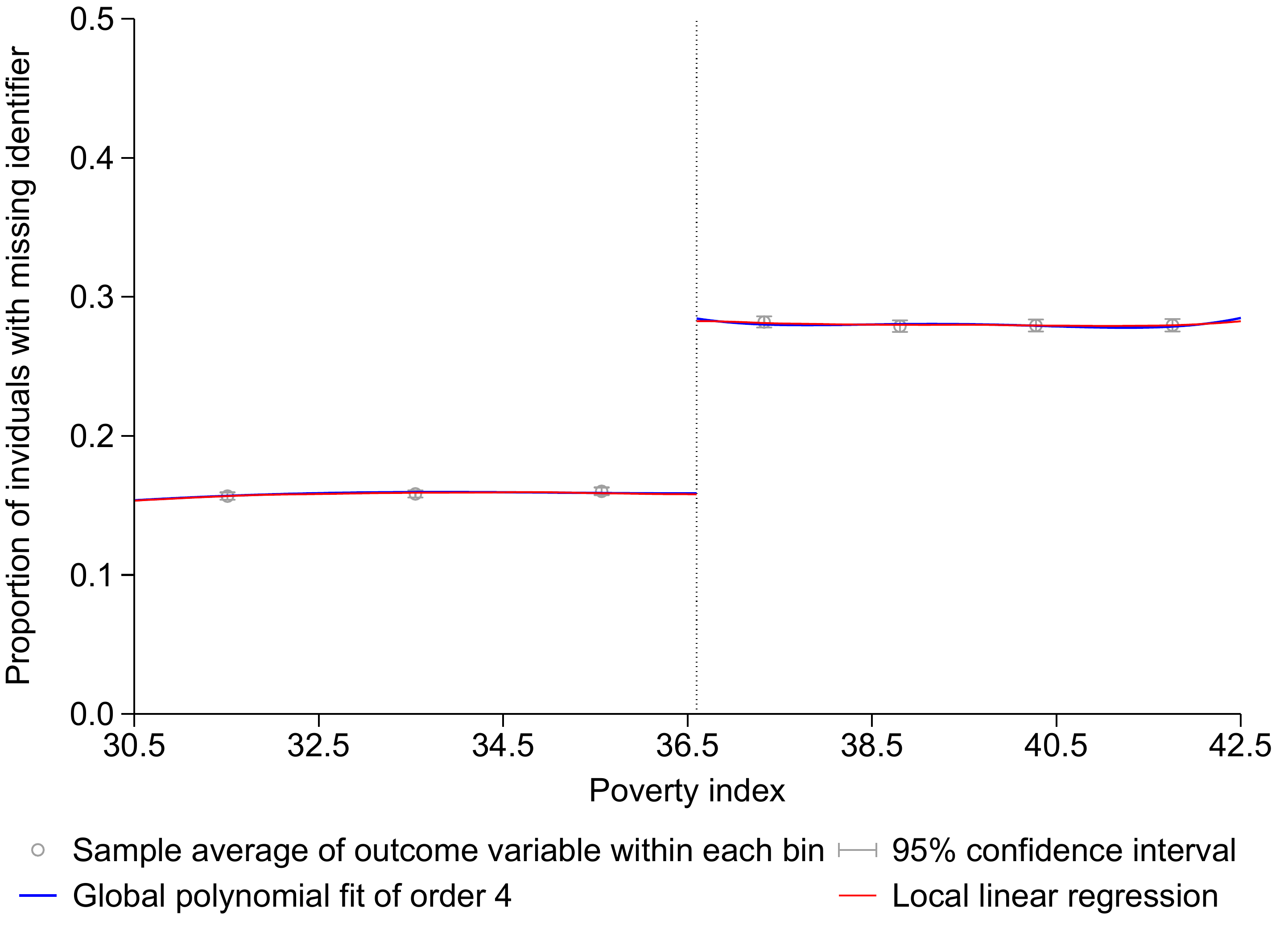} \\
	\end{subfigure}
	\begin{subfigure}[t]{1\textwidth}
		\centering
		\captionsetup{width=0.8\textwidth}
		\caption{Test for density discontinuity of the assignment variable including only individuals with identity card number}
		\includegraphics[width=0.7\textwidth, trim = 0cm 0cm 0cm 0cm, clip]{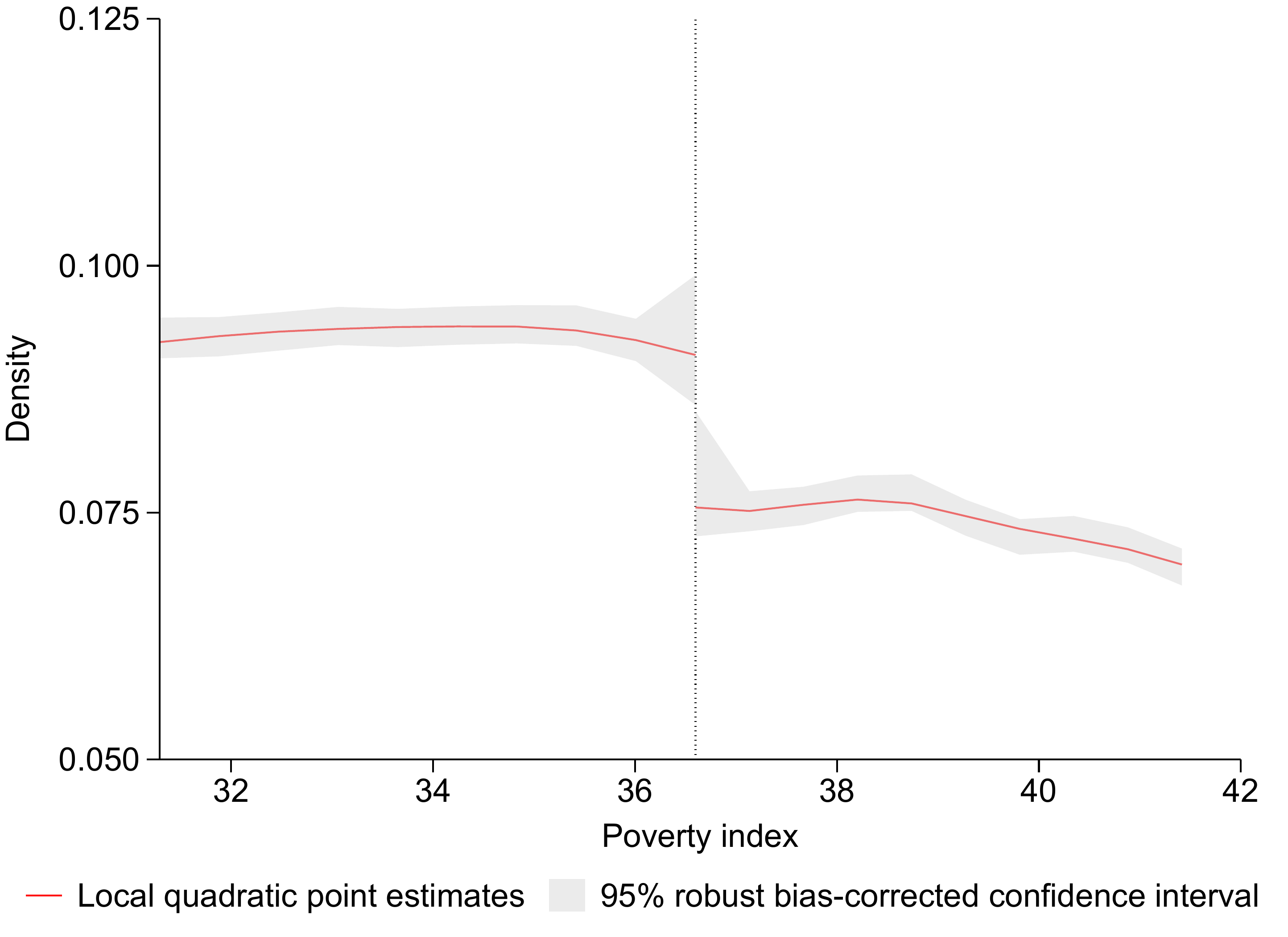} \\
	\end{subfigure}
	\begin{minipage}[c]{0.8\textwidth}	
		\justifying
		\noindent\textit{Notes}: The upper figure shows (i) the sample average of the outcome variable (proportion of individuals with missing identifier) within each bin and its 95\% confidence interval (grey circles), (ii) the fits of global fourth-order polynomial regressions estimated separately on either side of the cut-off (solid lines) and (iii) the fits of local linear regressions estimated separately on either side of the cut-off using a triangular kernel and data-driven MSE-optimal bandwidths. The construction of both plots uses the IMSE-optimal number of disjoint bins. The lower figure explores whether there is a discontinuity in the density of the poverty index using local quadratic regressions estimated separately on either side of the cut-off using a triangular kernel and data-driven MSE optimal bandwidths. The results of the test allow rejecting the hypothesis of continuity ($p-\text{value} = 0.000$).\\ 
		\noindent\textit{Source}: Authors' analysis from \textcite{socialregistry2025} and \textcite{socialsecurity2025}.
	\end{minipage}
\end{figure}

% Figure A3. ITT effects conditional on sector of activity
\begin{figure}[p]
	\footnotesize
	\captionsetup{width=0.80\textwidth}
	\caption{ITT effects on the probability of remaining employed in the formal sector month by month conditional on sector of activity in February 2020 for individuals living in rural areas in 2008/2009}
	\label{Figure A3}
	\centering 
	\includegraphics[trim={0 0 0 0}, width=0.80\textwidth]{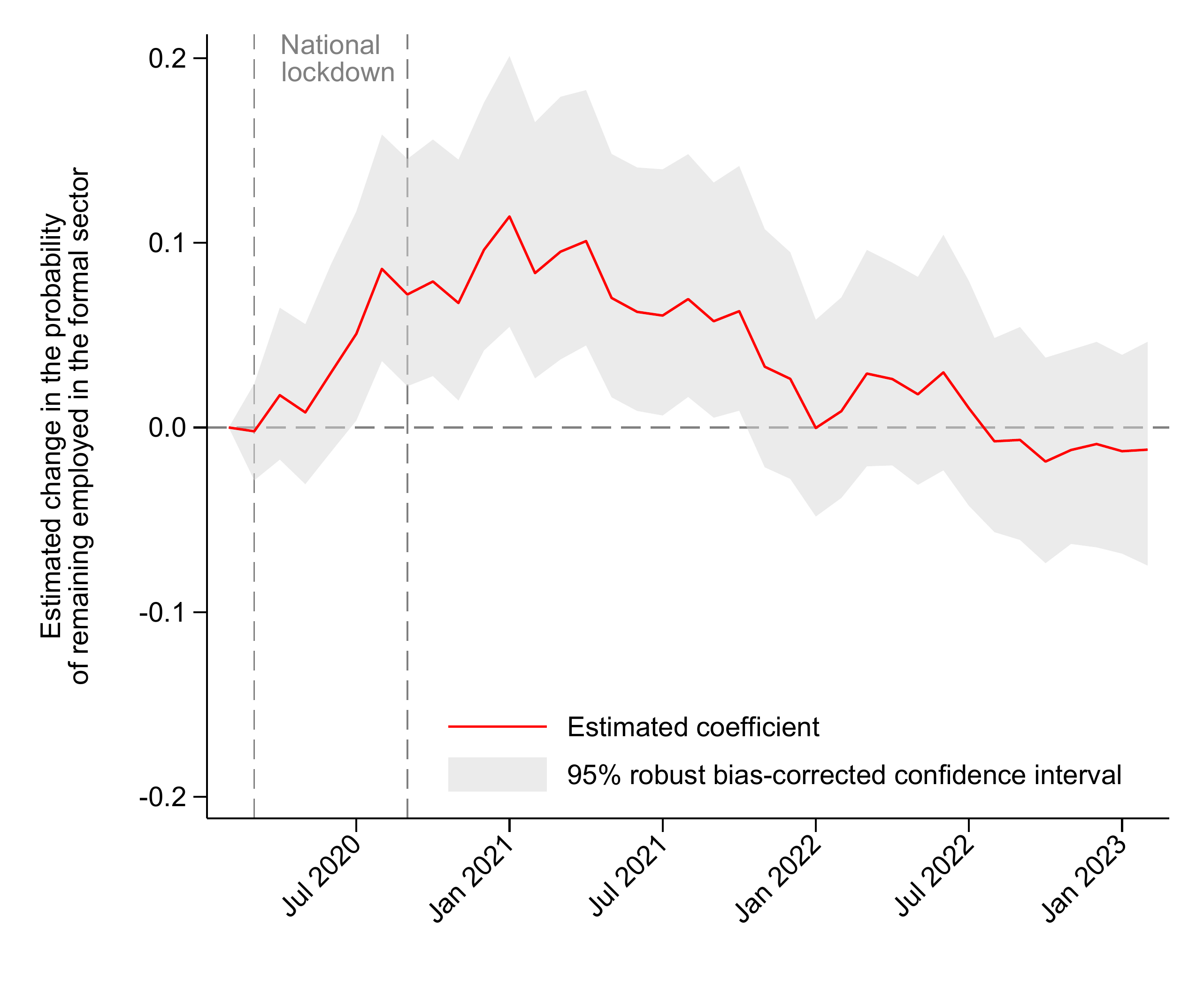}\\
	\noindent\makebox[\textwidth][c]{
		\noindent\begin{minipage}[c]{0.80\textwidth}	
			\justifying
			\noindent\textit{Notes}: The figure shows the estimated coefficients and the 95\% robust bias-corrected confidence intervals of the regression discontinuity exercises estimated separately for each month. Each analysis consists in local linear regressions (which allow for different slopes on either side of the cutoff) estimated using a triangular kernel and data-driven MSE-optimal bandwidths. All of them include the following covariates: child gender, child age, squared child age, child ethnicity, household head's educational attainment, household head's marital status, household size, area of residence (urban or rural), industry (in February 2020) fixed effects and canton fixed effects. Standard errors are clustered at the household level using robust inference.\\ 
			\noindent\textit{Source}: Authors' analysis from \textcite{socialregistry2025} and \textcite{socialsecurity2025}.
		\end{minipage}
	}
\end{figure}

% Figure A4. Effect heterogeneity in rural areas
\begin{figure}[p]
	\scriptsize
	\captionsetup{width=0.85\textwidth}
	\caption{ITT Effect heterogeneity for individuals in rural areas in 2008/2009}
	\label{Figure A4}
	\centering 
	\includegraphics[width=0.85\textwidth]{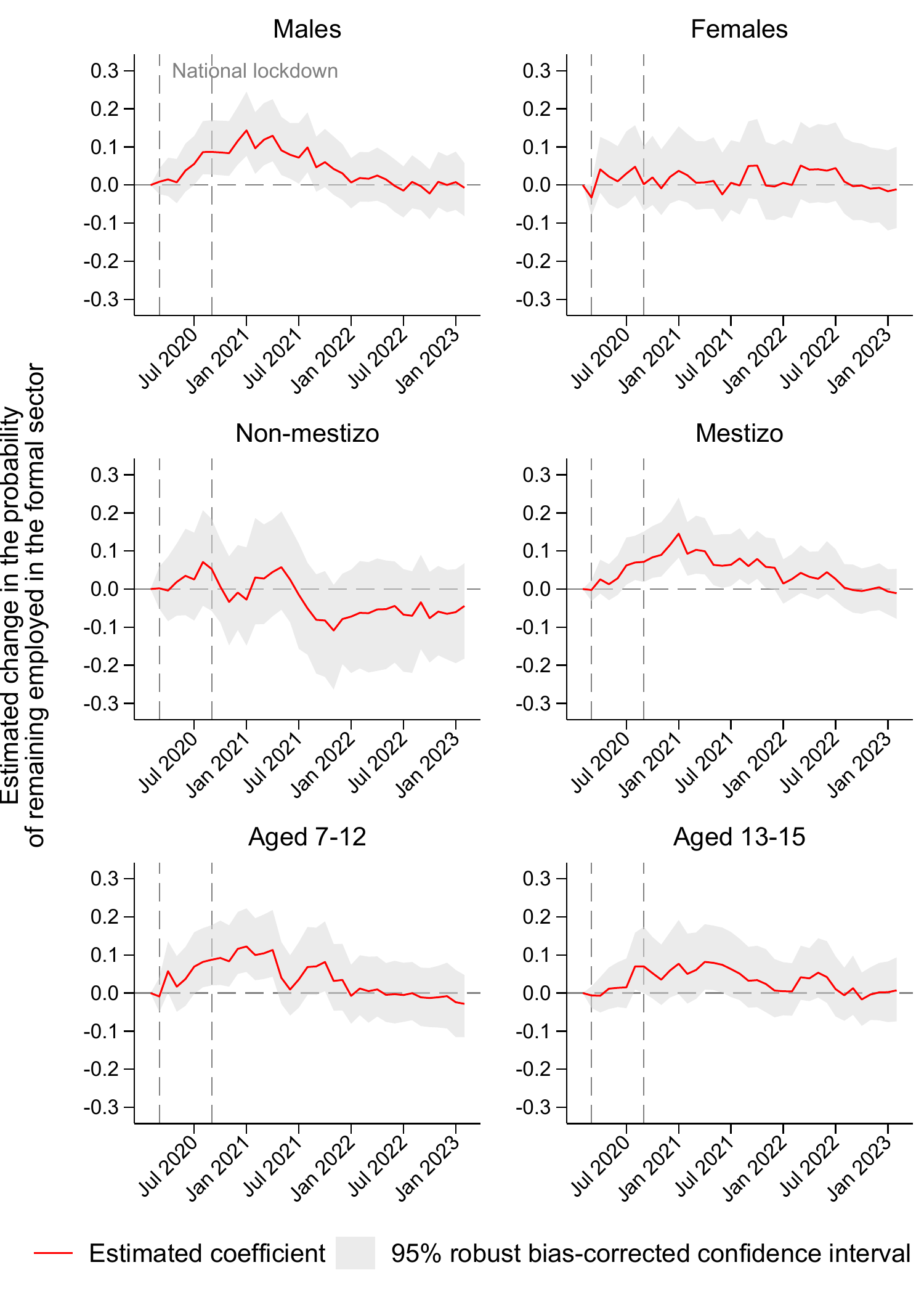}\\
	\noindent\makebox[\textwidth][c]{
		\noindent\begin{minipage}[c]{0.85\textwidth}	
			\justifying
			\noindent\textit{Notes}: The figures show the estimated coefficients and the 95\% robust bias-corrected confidence intervals of regression discontinuity exercises estimated separately for each month. Each analysis consists in local linear regressions (which allow for different slopes on either side of the cutoff) estimated using  triangular kernel and data-driven MSE-optimal bandwidths. All of them include the following covariates: child gender, child age, squared child age, child ethnicity, household head's educational attainment, household head's marital status, household size---excluding the covariate associated to each group appropriately---and canton fixed effects. Standard errors are clustered at the household level using robust inference.\\ 
			\noindent\textit{Source}: Authors' analysis from \textcite{socialregistry2025} and \textcite{socialsecurity2025}.
		\end{minipage}
	}
\end{figure}

% Figure A5. First stage of the fuzzy RDD in March 2020 for the whole sample and individuals living in rural areas in 2008/2008
\begin{figure}[p]
	\scriptsize
	\captionsetup{width=0.85\textwidth}
	\caption{First stage of the fuzzy RDD in March 2020 for the whole sample and for individuals living in rural areas in 2008/2009}
	\label{Figure A5}	
	\centering
	\medskip 
	\noindent\makebox[\textwidth][c]{
		\noindent\begin{minipage}[c]{0.85\textwidth}
			\centering	
			\includegraphics[width=0.85\textwidth]{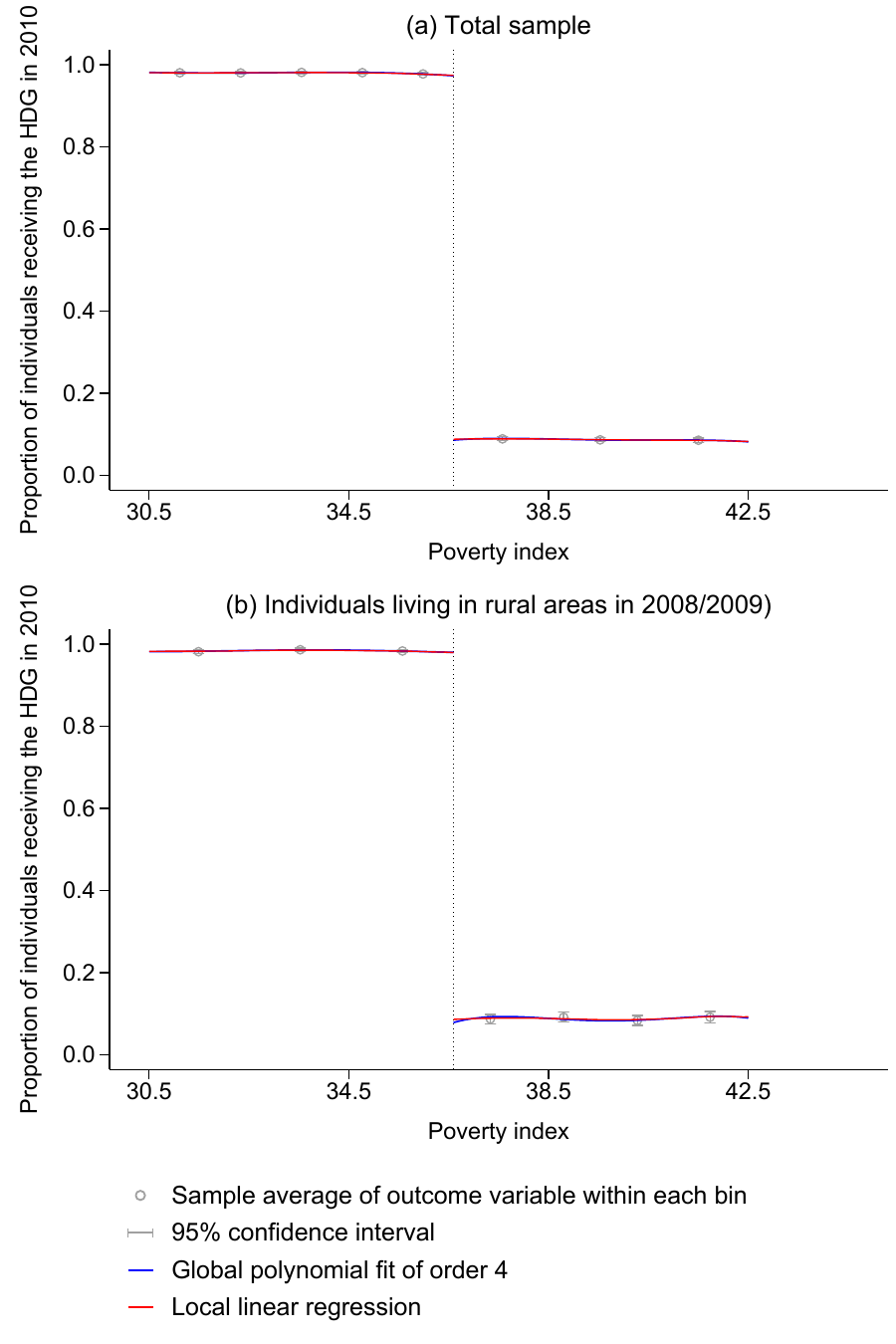} \\
			\justifying
			\noindent\textit{Notes}: The figures shows (i) the sample proportion of the total sample (upper panel) and non-mestizo individuals (lower panel) treated (receiving the HDF in 2010) within each bin and its 95\% confidence interval (grey circles), (ii) the fits of global fourth-order polynomial regressions estimated separately on either side of the cut-off (blue line) and (iii) the fits of local linear regressions estimated separately on either side of the cut-off using a triangular kernel and data-driven MSE-optimal bandwidths (red line). The construction of all plots uses the IMSE-optimal number of disjoint bins.\\ 
			\noindent\textit{Source}: Authors' analysis from \textcite{socialregistry2025} and \textcite{socialsecurity2025}.
		\end{minipage}
	}
\end{figure}

\end{document}